\documentclass[11pt]{article}
\usepackage{graphicx}% Include figure files
\usepackage{bm}% bold math
\usepackage{amsmath}
\usepackage{comment}
\usepackage{booktabs}
\usepackage{epstopdf}
\usepackage{amsfonts}
\usepackage{amssymb}
\usepackage{epsfig}
\usepackage{natbib}
\usepackage{subcaption}
\usepackage{setspace}
\usepackage{url}
\usepackage[hidelinks]{hyperref}
\usepackage[flushmargin]{footmisc}

\usepackage[all]{nowidow}

\usepackage[english]{isodate}

\hypersetup{
	colorlinks   = true, %Colours links instead of ugly boxes
	urlcolor     = black, %Colour for external hyperlinks
	linkcolor    = blue, %Colour of internal links
	citecolor   = blue %sepia %Colour of citations
} 
\usepackage{titlesec}
\titleformat*{\section}{\normalsize\bfseries}
\titleformat*{\subsection}{\normalsize\bfseries}
\titleformat*{\subsubsection}{\normalsize\bfseries}

\newtheorem{result}{\bf Result}

\newtheorem{conjecture}{\bf Conjecture}
\newtheorem{proposition}{\bf Proposition}

\begin{document}

\sloppy

\title{Evidence Gathering under Competitive and \\ Noncompetitive Rewards}
\author{Philip Brookins\thanks{University of South Carolina, Columbia SC, USA, \protect\url{philip.brookins@moore.sc.edu}, ORCID: 0000-0002-1959-9186.} \and Jennifer Brown\thanks{Corresponding author: University of Utah, Salt Lake City UT, USA, \protect\url{jen.brown@eccles.utah.edu}, ORCID: 0009-0002-4011-1939.} \and Dmitry Ryvkin\thanks{RMIT University, Melbourne VIC, Australia, \protect\url{d.ryvkin@gmail.com}, ORCID: 0000-0001-9314-5441}}

\date{\printdayoff\today}

\maketitle
\begin{abstract}
\noindent Reward schemes may affect not only agents' effort, but also their incentives to gather information to reduce the riskiness of the productive activity. In a laboratory experiment using a novel task, we find that the relationship between incentives and evidence gathering depends critically on the availability of information about peers' strategies and outcomes. When no peer information is available, competitive rewards can be associated with more evidence gathering than noncompetitive rewards. In contrast, when decision-makers know what or how their peers are doing, competitive rewards schemes are associated with less active evidence gathering than noncompetitive schemes.  The nature of the feedback---whether subjects receive information about peers' strategies, outcomes, or both---also affects subjects' incentives to engage in evidence gathering. Specifically, only combined feedback about peers' strategies and performance---from which subjects may assess the overall relationship between evidence gathering, riskiness, and success---is associated with less evidence gathering when rewards are based on relative performance; we find no similar effect for noncompetitive rewards. \\



\end{abstract}

\newpage
\onehalfspacing

\section{Introduction}

In many settings, individuals make choices about the riskiness of their actions by deciding how much evidence to gather. For example, a salesperson might exert effort to close a deal, hustling to satisfy the customer while over-extending delivery promises without checking their inventory. Financial analysts might neglect time-consuming research in favor of producing more reports. A student might decide not to devote time to studying course material in favor of other more enjoyable pursuits, hoping to get lucky with familiar questions on the final exam. In each case, individuals have the opportunity to gather evidence---about inventory, historical market performance, or concepts from class---to reduce the riskiness of their actions. However, evidence gathering is costly, and the incentives to engage in this risk-reducing activity depends on the nature of the risk-reward trade-off.

Reward structures vary considerably by setting: Salespeople may be paid by the hour, by commission, or according to a tournament-style scheme; financial analysts may be low-wage interns, salaried employees, or associates facing high-powered incentives; and students may be assigned credit/no-credit, scored on a set scale, or graded on a curve. Each reward structure influences participants' incentives to invest in costly risk-reducing evidence gathering. Of course, rewards schemes are never imposed in a vacuum; recently, there has been an increasing focus on the impact of social influences on economic decisions.\footnote{Although psychologists have a long history of research on social influence, only recently has economic theory modeled social influence as it relates to settings with risk. For example, \cite{Maccheroni-et-al:2012} present an axiomatic foundation for interdependent preferences that supports the claim that the observation of peers' outcomes can be useful in learning to improve one's own choices.} More specifically, individuals' decisions to gather evidence may be influenced by social and informational factors, such as their relative and absolute performances to date, the norms around evidence gathering within their organizations and industry, and the information available to them before and during the competition. 

On the other side of the market, administrators, managers, and teachers may differ in their perspective on how much evidence they would want to be gathered before the risky action is taken. For example, the financial firm may prefer more superficial reports over fewer detailed ones. In contrast, teachers may prefer that students study a small number of chapters in depth rather than skimming over the whole textbook (and they likely prefer both of these study techniques over no studying at all). Understanding what compensation schemes and peer feedback structures induce desirable evidence gathering can inform managers', administrators', and teachers' design choices.

In this paper, we consider the interaction between incentives and peer feedback and examine, first, how the presence and nature of feedback about peers affects individuals' incentives to gather evidence to reduce the riskiness of their choices. More precisely, we ask: How is an individual's incentive to gather risk-reducing evidence influenced by the availability of feedback about the performance and/or strategies of his or her peers? And how does the relationship between peer feedback and evidence gathering change when individuals are rewarded based on their relative versus absolute performance? 

We use an incentivized forecasting task in a controlled laboratory environment to answer these questions. Field data on individuals' incentives, evidence gathering strategies, and peer feedback are largely unavailable and, even if available, they would be difficult to analyze due to inherent endogeneity. The laboratory allows us to use the random assignment of subjects to treatments to infer causal relationships between peer feedback, incentives, and individuals' decisions to engage in risk-reducing evidence gathering.

The experimental environment is novel and was designed to approximate a real-world setting where a trade-off between risk and return is generated via evidence gathering under a time (or budget) constraint \citep[for a related setting see, e.g.,][]{Page-Siemroth:2017}. It is motivated with a hypothetical, framed scenario that asks subjects to imagine themselves as financial analysts making simple predictions about future stock performance. Specifically, analysts can gather evidence prior to announcing their predictions, but that research is costly and limits the resources available for future forecasts. The experimental task asks subjects to decide how much evidence to gather before submitting a binary prediction. The analyst faces a trade-off between the volume of forecasts that he or she can complete and their accuracy. Depending on treatment, the analyst is compensated according to an absolute or relative performance-based pay scheme and receives information about the evidence gathering strategies and/or forecasting outcomes of his or her peers.

Overall, our results suggest that the relationship between rewards and evidence gathering depends critically on features of the environment, including the nature of the task and the availability of feedback about peers' previous strategies and outcomes.\footnote{We frame the results as decision makers' responses to observing (or not) the strategies and outcomes of other competitors. Anticipation of being observed by others may also influence individuals' attitudes to risk \citep{Weigold-Schlenker:1991} and other actions.}  More specifically, when no peer feedback is available, our relative rewards scheme is associated with more effort to gather risk-reducing evidence than noncompetitive rewards. The presence of feedback reverses this relationship: when competitors learn about their peers' actions and performance, the relative incentive scheme is associated with less evidence gathering than a noncompetitive scheme.

The nature of the peer-related feedback also matters. In a noncompetitive rewards setting, subjects engage in more evidence gathering when exposed to simple feedback about either peers' strategies or outcomes, yet combined feedback about strategies and outcomes has no effect. The opposite is true for settings with competitive rewards: simple information has little impact, but subjects engage in less evidence gathering when they receive feedback about both the strategies and scores of their peers.  We also find some degree of sensitivity to relative position under competitive incentives when the peer feedback is available, but no similar effects in settings with noncompetitive rewards. Although previous studies have documented a preference for conformity in settings with risk \citep{Goeree-Yariv:2015}, we find no evidence that subjects' strategies converge over time within a group or that subjects' imitate their peers' strategies.  

Our paper contributes to two streams of the literature on incentives: the first aims to understand the impact of specific compensation systems on risk-reducing evidence gathering and the second explores the impact of feedback on performance. We find that decision-makers' risk taking is influenced by both of these features---incentives to gather risk-reducing evidence depends critically on both the compensation style and the availability of feedback about peer performances.  

Compensation structure and performance feedback are both important dimensions of organizational design. The implication of our results varies by context. For example, when risk taking stimulates creativity and innovation, firms may wish to pair highly competitive rewards schemes with rich feedback about peers' strategies and outcomes. When risk taking undermines the stability of a firm or industry, firms using relative compensation schemes may wish to restrict information about peers' actions and outcomes. If peer feedback is readily available, then firms may wish to rely on noncompetitive rewards to encourage evidence gathering that reduces undesirable risk taking.  While a firm's overall performance likely hinges on a variety of factors in addition to its workers' effort and risk taking, our research suggests that firms should pay particular attention to the nuanced relationship between incentives, peer feedback, and individuals' decisions to engage in risk-reducing evidence gathering.

The paper is organized as follows: Section \ref{sec:lit} highlights the current paper's position in the literature on risk taking, peer feedback, and incentives; Section \ref{sec:design} describes our novel experimental design; Section \ref{sec:predictions} provides a theoretical perspective on the experimental task and describes specific empirical predictions; Section \ref{sec:results} summarizes and interprets the results of the experiment; and Section \ref{sec:conclusion} concludes with a discussion of the paper's implications for understanding the relationships between evidence gathering, risk taking, and peer feedback in competitive and noncompetitive settings.

\section{Related literature}  
\label{sec:lit}

To our knowledge, there is little existing work---theoretical, experimental, or empirical---on how individuals' incentive to gather risk-reducing evidence is affected by their competitive environment, including the compensation scheme and availability of peer feedback. In the following section, we highlight our contribution by placing our work in the existing literatures on compensation schemes and risk taking; peer feedback and risk taking; and risk taking in non-strategic environments. %We also briefly contrast our research with the extant legal literature, a very different perspective on an ``evidence'' related question. 

%\textcolor{red}{DR: I've added a paragraph below. It seems to fit organically with the rest, not in contrast.} 

Although much of the theoretical literature comparing competitive and noncompetitive compensation schemes has focused on differences in the incentives for effort \cite[cf.][]{Lazear-Rosen:1981}, researchers have also modeled the incentives for risk taking under different types of pay schemes. In many existing models, tournament-style rewards induce more risk taking than comparable noncompetitive, piece rate rewards. For example, competitors in winner-take-all tournaments are predicted to choose maximal risk and zero effort, regardless of the prize spread \citep{Hvide:2002}. Similarly, in multi-prize settings, more risk taking is expected when prizes are awarded to a lower proportion of participants \citep{Gaba-et-al:2004}. 

Ability or interim position in the tournament (if it is revealed) may also affect risk taking. The common wisdom is that leaders tend to ``play it safe'' to preserve their position while followers take more risk trying to catch up: less able chicken producers adopt higher variance strategies \citep{Knoeber-Thurman:1994}, and mutual fund managers increase the riskiness of their portfolios when their mid-year performance is below the industry average \citep{Brown-et-al:1996}. The same regularities are observed in laboratory experiments by \cite{Eriksen-Kvaloy:2014} and \cite{Kirchler-et-al:2018}. In an asset trading experiment, larger bubbles are observed when subjects receive feedback about the performance of the top trader in their group \citep{Schoenberg-Haruvy:2012}. In a laboratory experiment on investment portfolio choice, when information on subjects' relative positions is available, leaders adjust their portfolios in the direction of negatively skewed assets, whereas followers prefer positively skewed assets \citep{Dijk-et-al:2014}. The result holds for both competitive and noncompetitive rewards, which suggests that ``playing it safe'' and ``catching up'' are driven mainly by social rather than monetary incentives. However, the relationship between ability (or relative position) and risk taking is not always negative.  In contrast to the standard theory, \cite{Taylor:2003} finds that leading mutual fund managers take more risk in their portfolio choices in the presence of interim reviews, whereas trailing managers take less risk. More generally, when players choose sequentially the riskiness of their production technology and their effort, adoption of the risky technology depends critically on the players' relative abilities, the incentives for effort, and their respective likelihoods of success \citep{Krakel:2008}. 

Evidence gathering strategies are discussed extensively in the legal literature, for example, in the context of admissibility of evidence across jurisdictions \citep[e.g.,][]{Daniele-et-al:2018}, and more generally, in terms of how they affect the likelihood of convictions \citep[e.g.,][]{Ling-et-al:2021}. Excessive risk taking, in the form of insufficient diligence in evidence gathering and processing by personnel, has been identified as a major cause of wrongful convictions in the US \citep{Morgan:2023}.

The early experimental literature comparing competitive and noncompetitive reward schemes focuses primarily on differences in effort.\footnote{In their groundbreaking paper, \cite{Bull-et-al:1987} note that rank-order tournaments are associated with increased variance in subjects' effort choices, which provides indirect evidence for differences in risk taking. Subsequent studies replicate these results. For a recent survey of the experimental literature on tournaments, see \cite{Dechenaux-et-al:2015}.} More recently, a number of experiments test the theoretical predictions about risk taking in tournaments. \cite{Nieken:2010} finds that, as predicted, subjects choose lower efforts when more noise is present in the tournament; however, contrary to the predictions, subjects fail to select the highest level of risk. \cite{James-Isaac:2000} and \cite{Robin-et-al:2021} observe a more intense formation of bubbles in an experimental trading market with tournament-style incentives. \cite{Vandegrift-Brown:2003} explore the role of ability differences and task difficulty in risk taking with tournament incentives and find that low-ability subjects are more likely to choose high-risk strategies, but only in a simpler task. \cite{Nieken-Sliwka:2010} find that the relationship between ability and risk taking may be more complex in the presence of correlated shocks.

It has been long understood that information about competitors' own and peer performance affects people's behavior in games \citep[e.g.,][]{Duffy-Feltovich:1999}.\footnote{In contrast, \cite{Eriksson-et-al:2009} vary the frequency of feedback about subjects' relative positions and find that feedback has no effect on effort in either noncompetitive or tournament environments.} Moreover, peer effects on effort have been identified in both strategic and nonstrategic settings. For example, \cite{Lount-Wilk:2014} report that feedback can mitigate the free-riding problem in group production: subjects work better in a group than alone when feedback on performance is provided and worse in a group than alone when it is not provided. \cite{Falk-Ichino:2006} examine peer effects in a real-effort individual production task.

Several studies explore the role of information and peer effects in subjects' risk-taking decisions in nonstrategic environments. \cite{Linde-Sonnemans:2012} find that subjects take less risk when they earn as much as a peer and more risk when they earn at least as much, which is the opposite to what is predicted by the prospect theory with a social reference point \citep[see also][]{Bault-et-al:2008}. \cite{Gamba-et-al:2017} find that when subjects observe their peers' wages, both wage leaders and wage followers take more risk in a subsequent task. \cite{Fafchamps-et-al:2015}, \cite{Karakostas-et-al:2021}, and \cite{Celse-et-al:2021} explore the relationship between risk-raking and information about peers' behavior in a set of similar experiments using an investment task
introduced by \cite{Gneezy-Potters:1997}, where an agent decides what part of an endowment $E>0$, $x\in[0,E]$, to invest into a lottery $L=(p,1-p;\alpha x,0)$, where $p$ is some probability and $\alpha>1$ is a parameter. Specifically, \cite{Fafchamps-et-al:2015} disclose both peers' investment decisions and realized earnings and document behavior consistent with a ``keeping up with the winner'' effect, as opposed to a ``keeping up with the Joneses'' effect; that is, subjects behave as if they are engaged in competition and increase investments so as to try and keep up with high-earning subjects. \cite{Karakostas-et-al:2021} vary whether or not peers' investment decisions are disclosed and find that disclosure leads to investment clustering at the group level, but the specific psychological mechanism behind the result is unexplored. To discriminate between mechanisms underlying risk taking in the presence of information about peers' investment decisions and earnings, \cite{Celse-et-al:2021} conduct a ``horse race'' between several models combining social preferences---such as outcome and action driven preferences, inequality aversion, conformism, and social loss aversion---and risk preferences with a CRRA utility function. Overall, they find that investment behavior is best predicted by competitive preference models, suggesting that individuals derive utility from favorable comparisons of own investments and earnings to those of their peers. \cite{Lahno-Serra-Garcia:2015} separate the effects of imitation of choices from relative payoff considerations and show that peers' choices have a significant impact on risk taking. \cite{Cooper-Rege:2011} suggest that such imitation is driven by a ``social interaction effect.'' That is, a person's utility from taking an action increases if others take the same action; social regret explains data better than preference for conformity. \cite{Eriksen-Kvaloy:2017} explore the effects of competitiveness on risk taking in a tournament where it is optimal (weakly dominant) for subjects to take no risk. More risk taking is observed when the number of players in the tournament increases and when feedback on the winner's earnings in the previous round is provided. 

We are aware of only one study that explicitly compares risk taking under competitive and noncompetitive pay schemes under different information conditions, although it focuses on different features of the environment. \cite{Eriksen-Kvaloy:2014} use a simple lottery investment task experiment in which feedback on both strategies and outcomes is provided at various frequencies. Consistent with theory, more frequent feedback leads to less risk taking under a noncompetitive compensation scheme, but more risk taking when rewards are competitive. In contrast to these authors' interest in the frequency of feedback, we focus on the effect of different types of information contained within the feedback.

\section{Experimental design and procedures} 
\label{sec:design}

We conducted a laboratory experiment to study the effect of information about peers' strategies and performance on subjects' risk-reducing evidence gathering in competitive and noncompetitive settings. To better explain the experimental task, we motivated subjects with a hypothetical, framed scenario by asking them to imagine themselves as financial analysts making projections about the future performance of particular stocks.\footnote{Although we frame the experiment with this hypothetical choice faced by financial analysts, many other professionals face this type of trade-off in their daily tasks. For example, a journalist might have to decide whether to print more sensational, but possibly speculative, stories or take additional time to secure more sources to verify each story; and a detective investigating crimes can collect more evidence to build more convincing cases, or close more cases with less evidence risking unfavorable court rulings.} 
Specifically, the analyst must assess whether the future price will be higher or lower than the current price, and the analyst's pay reflects both the volume and accuracy of the forecasts. Forecasts are based on information gathered about the stocks being considered, but gathering information is costly. The scenario highlights the analyst's trade-off: More information improves accuracy but reduces volume (see section \textquotedblleft The scenario\textquotedblright\ of the experimental instructions in \ref{sec:instructions}).\footnote{\cite{Vandegrift-et-al:2007} use a different forecasting task to study performance when subjects can choose between relative (competitive) and absolute (noncompetitive) performance-based incentive schemes. Unlike in our design, their subjects could not choose explicitly how much information to gather in the face of a risky decision.}

Each experimental session consisted of two parts. First, subjects' risk aversion and ambiguity aversion were assessed using list elicitation methods similar to those described in \cite{Sutter-et-al:2013}. During each assessment, subjects were presented with a list of 20 choices between earning \$2.00 for correctly guessing the color of a ball drawn randomly from an urn and a sure amount of money. The sure amounts of money increased from \$0.10 to \$2.00, and subjects were asked to choose the point at which they were willing to switch from the draw to the sure amount. For the risk-aversion assessment, subjects were informed that the urn contained 10 green balls and 10 red balls. For the ambiguity aversion assessment, subjects were informed that the urn contained balls of the two colors, but the exact number of balls of each color was not disclosed.\footnote{Unbeknownst to subjects, the share of red balls was generated randomly from a uniform distribution.} The results and payoffs from this part of the experiment were withheld until the end of the session.

The second part of the experiment consisted of a forecasting game. Subjects participated in several periods of play, divided into blocks. At the beginning of each period, a subject was presented with an image of 15 blank cards on his or her computer screen. When flipped over, each card was either green or red. The color of the card was determined randomly and either color was equally likely to appear. A subject's task was to predict whether the majority of the 15 cards was green or red. The subject started by choosing how much information to collect, by deciding how many cards---between 5 and 15---to flip at once.\footnote{The five card minimum on the number of flips served two purposes. First, it allowed subjects to make sufficiently many forecasts, but not too many, to avoid an excessively long experiment. Second, the dependence of expected payoffs on decisions is weak within the available strategy space, which makes the environment suitable to study the effects of feedback.} Having observed the cards, the subject then made his or her prediction (examples of decision screens are provided in  \ref{sec:screens}). Subjects could gather less risk-reducing evidence by flipping fewer cards---a higher-risk strategy, as the subject has less evidence on which to base his or her assessment. The highest-risk strategy involves flipping only 5 cards; in contrast, the lowest-risk strategy is one in which a subject reveals all 15 cards and, therefore, can infer a correct assessment for sure. Once the assessment had been submitted, all of the cards were revealed to the subject, and he or she was told if his or her forecast was correct or incorrect.

Periods were divided into blocks in which a subject was allowed to flip a total of 100 cards; a counter on the screen displayed the number of remaining flips. A subject repeated the same assessment task---forecasting the majority color---until he or she had exhausted all 100 flips. Subjects were not constrained to follow the same strategy in each period of a block; as a result, a subject could make between 7 and 20 assessments in a given block.\footnote{Recall that subjects could not flip fewer than five cards in each period and, for that reason, towards the end of the block they would not be allowed to flip a number of cards such that fewer than five cards remained.}

A subject's score in a block was calculated as the number of correct assessments minus the number of incorrect assessments. At the end of each block, subjects were given a complete history of their individual forecasts, including the number of green and red cards flipped each period, their majority color assessment, and whether the assessment was correct or incorrect. Additionally, subjects received a summary of their own performance (i.e. their score for the block and their average information gathering strategy, measured by the average number of cards flipped per period). Depending on the experimental treatment, subjects were also presented with information about other participants' strategies and/or scores (a sample feedback screen is provided in  \ref{sec:screens}).

At the beginning of the forecasting game, subjects were randomly assigned to groups of five participants. The identities of group members were not revealed to participants and were described only by identification numbers 1 through 5. Groups and subject identification numbers remained the same throughout the experiment.

We implemented two reward schemes: \textit{noncompetitive} and \textit{competitive}. Under the noncompetitive scheme, a subject's payoff in a block was calculated as \$1.50 multiplied by his or her score for the block. Under the competitive scheme, a subject's payoff in a block was calculated according to the subject's rank by score in his or her group, with ties broken randomly. The top ranked subject in the group earned \$2.50 multiplied by his or her score; second, third and fourth ranked subjects earned \$1.50 multiplied by their individual scores; and the subject in fifth place earned \$0.50 multiplied by his or her score.\footnote{The presence of three distinct piece rates intensifies competition because subjects may not only compete to earn the top position but also to avoid the bottom, which has been shown experimentally to lead to higher average effort than comparable two-prize schemes \citep{Dutcher-et-al:2015, Gill-et-al:2018}.}

We also implemented four peer information conditions: (i) no feedback about peers' evidence gathering strategies or scores; (ii) feedback about peers' evidence gathering strategies only, shown as the average number of cards flipped in the previous block for each group member; (iii) feedback about peers' scores only, shown as the score in the previous block for each group member; and (iv) feedback about both peers' strategies and scores.

The resulting eight treatments (2 reward schemes$\times$4 feedback conditions) were conducted following a between-subject design. All subjects in a given session were in the same treatment. The number of subjects and groups in each treatment are summarized in Table \ref{tab:treatments}.

At the beginning, all subjects received instructions for the noncompetitive scheme only. In the first block of periods, identical across all treatments, subjects were rewarded according to the noncompetitive scheme. After the first block, the experiment was paused. All subjects received information about their own performance and, in addition, depending on the treatment they were in, information about their peers' strategies, scores, or both, from the first block. Subjects in treatments with noncompetitive rewards were informed that the following blocks would continue under the same incentives. In treatments with competitive rewards, subjects were told that starting from the second block incentives would change, and provided instructions on the competitive rewards. 

\begin{table}[tbp]
\begin{center}
\caption{Summary of experimental treatments.}
\begin{tabular}{lcccc}
\hline\hline

				& \multicolumn{2}{c}{\textbf{Noncompetitive}} & \multicolumn{2}{c}{\textbf{Competitive}} \\
\hline
& & & &  \\ 
			& 	&	& \multicolumn{2}{c}{\$2.50 $\times$ Score if rank=1}			 \\
Rewards			&\multicolumn{2}{c}{\$1.50 $\times$ Score}		& \multicolumn{2}{c}{\$1.50 $\times$ Score if rank=2,3,4}\\
		&	&	& \multicolumn{2}{c}{\$0.50 $\times$ Score if rank=5}\\
& & & & \\ 
\hline	
 \textbf{Peer information} & \textbf{\# of subjects} & \textbf{\# of groups} &\textbf{ \# of subjects} & \textbf{\# of groups} \\
\hline
None							&	45 & -	& 45 & -	\\	
Strategies 		&	35 & 7 	& 40 & 8 	\\
Scores							&	40 & 8	& 40 & 8 	\\
Both		&	75 & 15 & 80 & 16 	\\
\hline\hline
\end{tabular}
\label{tab:treatments}
\end{center}
\end{table}

Subjects played a total of four blocks---the first block identical across treatments and blocks 2, 3 and 4 under the treatment conditions---with feedback repeated according to treatment after the second and third blocks. At the end of the session, actual payments were based on subjects' payoffs from one randomly selected block.

At the end of each session, participants were asked the following open-ended questions about their strategies in the forecasting game: (i) What was your strategy? (ii) Did your strategy change over time, and if so, how? (iii) Did your strategy change when you learned what others in your group were doing, and if so, how? and (iv) Did you follow/imitate anyone else's strategy? If so, whose strategy? Questions (i) and (ii) were asked in all treatments and questions (iii) and (iv) were asked only in treatments in which subjects received feedback about their peers' strategies and/or outcomes. Answers to the four questions were categorized by two independent reviewers using a common rubric with binary coding.\footnote{The complete rubric is available from the authors by request.} Responses from the two reviewers were aggregated by taking the minimum of their codings (i.e., an answer was coded as belonging to a category only if it was assigned to that category by both reviewers). 

Four hundred subjects, 49\% of whom are female, were recruited using ORSEE \citep{Greiner:2015} from the pool of more than 3,000 Florida State University students who preregistered for participation in experiments at the XS/FS lab. Each subject participated in one session. We conducted 20 sessions of the experiment---four sessions for each of the treatments with peer information about both evidence gathering strategies and scores, and two sessions otherwise. Sessions lasted approximately 90 minutes, including instructions and payment. On average, subjects earned \$22.60, including a \$10 participation payment.\footnote{It was possible for a subject to earn less than the show-up payment provided the subject's score in the randomly chosen block in the forecasting game was sufficiently negative. However, subjects were also paid for the risk and ambiguity tasks, and (luckily) total earnings ended up being above \$10 in all cases.} The experiment was implemented with the software package z-Tree \citep{Fischbacher:2007}.

\section{Theory and conjectures}
\label{sec:predictions}

Before describing the results in Section \ref{sec:results}, we provide a brief theoretical description of the experimental task and formulate specific conjectures. Our empirical findings suggest that subjects' choices deviate significantly from the theory predictions; even so, the framework provides helpful structure to understanding risk taking in the forecasting game.\footnote{We have no evidence that these deviations are a result of misunderstanding or confusion about the experimental task. In fact, we have confidence in the ecological validity of our framing. Although the experiment involved several stages and required subjects to consider a non-trivial risk-return trade-off, the experiment was straightforward to explain and the underlying decisions were familiar ones. After all, the costly evidence gathering decision, in the face of a volume vs. reliability trade-off, is one that university students face regularly in their campus lives. For example, when completing a lengthy timed exam, students must decide between answering more questions or spending more time on each question to increase the probability of answering it correctly. Moreover, none of our subjects mentioned confusion or misunderstanding in the post-experimental questionnaire.}

\subsection{Theoretical predictions}
\label{sec:theorypredictions}

Consider first the decisions that subjects face in the noncompetitive environment. Let $M$ denote the (odd) number of cards considered in each period, and let $N$ denote the total number of cards that can be flipped in one block (e.g. in our experiment, $M=15$ and $N=100$). 

Suppose in some period a subjects flips $n$ cards out of $M$ and uses the majority color among the flipped cards to construct his or her forecast. Let $r$ denote the number of red cards among the $n$ flipped cards. Suppose $n$ is odd and $r>n-r$ (i.e., red is the predicted majority color). The probability that this forecast is correct is
\begin{equation}
	\label{pnr}
	p_{n,r}= \left\{\begin{array}{ll}\left(\frac{1}{2}\right)^{M-n}\sum_{m=\frac{M+1}{2}}^{M-n+r}\binom{M-n}{m-r}, & r<\frac{M+1}{2}\\ 1, & {\rm otherwise}\end{array}\right.
\end{equation}
Indeed, the forecast will be correct with probability one if $r\ge \frac{M+1}{2}$ (i.e., if flipped red cards constitute the majority of $M$ cards). Otherwise, the probability of a correct forecast is given by the sum over all possible realizations of the total number of red cards (both opened and not), $m$, such that red is the majority color.\footnote{The probability of obtaining each configuration of unopened cards is $(\frac{1}{2})^{M-n}$, and the number of possible configurations of unopened cards for a given number of red cards $m$ is $\binom{M-n}{m-r}$.} The probability of a correct forecast after flipping an odd number of cards $n$ is, therefore,
\begin{eqnarray}
	\label{pn_odd}
	p_n = 2\left(\frac{1}{2}\right)^n\sum_{r=\frac{n+1}{2}}^n\binom{n}{r}p_{n,r}, \quad n {\rm \ odd}.
\end{eqnarray}
In equation (\ref{pn_odd}), the factor 2 arises because there are two possible majority colors; $(\frac{1}{2})^n$ is the probability of each realization of colors of $n$ cards; $\binom{n}{r}$ is the number of such realizations for a given $r$; and the summation includes all cases in which one color is the majority among the flipped $n$ cards. 

Suppose now that $n$ is even. For $r>\frac{n}{2}$ (i.e., when $r$ is the majority color), $p_{n,r}$ is given by equation (\ref{pnr}). For $r=\frac{n}{2}$ (i.e., with probability $\left(\frac{1}{2}\right)^n\binom{n}{\frac{n}{2}}$), the probability of a correct forecast is $\frac{1}{2}$. Therefore, the probability of a correct forecast after flipping an even number of cards $n$ is
\begin{eqnarray}
	\label{pn_even}
	p_n = \frac{1}{2}\left(\frac{1}{2}\right)^n\binom{n}{\frac{n}{2}}+ 2\left(\frac{1}{2}\right)^n\sum_{r=\frac{n+2}{2}}^n\binom{n}{r}p_{n,r}, \quad n {\rm \ even}.
\end{eqnarray}

\begin{figure}[tbp]
\begin{center}
{\includegraphics[width=3.0in]{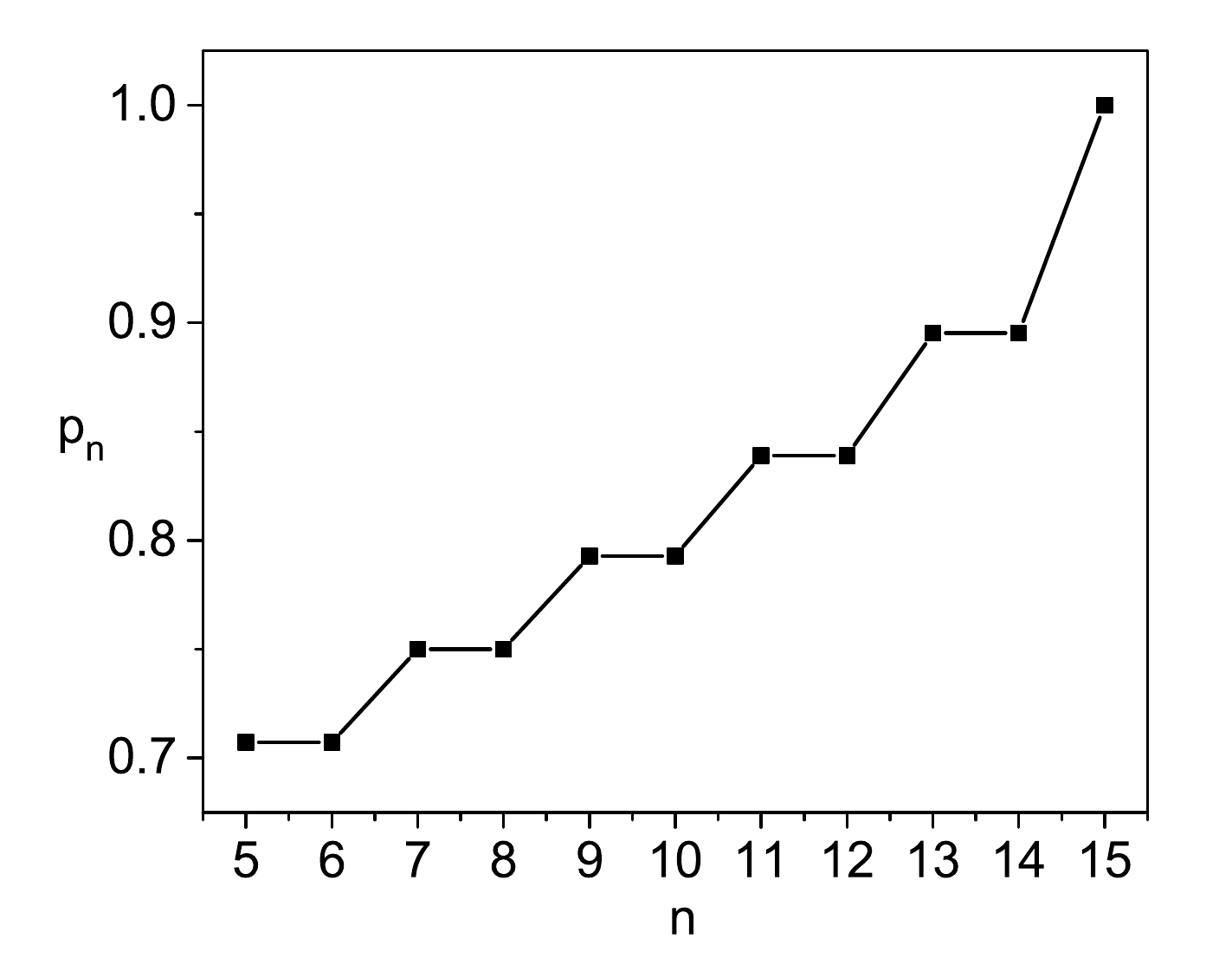}}
\end{center}
\caption{The probability, $p_n$, of correctly guessing the majority color of $M=15$ cards after flipping $n$ cards. The points are obtained using Equations (\ref{pn_odd}) and (\ref{pn_even}), and connected by straight lines for better visualization.}
\label{fig:pn}
\end{figure}

The resulting probability of a correct forecast, $p_n$, is shown in Figure \ref{fig:pn}. The first observation from the figure is that it is never optimal to flip an even number of cards $n$ because the same probability of a correct forecast can be reached by flipping $n-1$ cards. Indeed, when an even number of cards $n$ is flipped, the numbers of revealed red and green cards can either be equal or differ by at least two. In the former case, the probability of a correct guess is $\frac{1}{2}$; therefore, by flipping $n-1$ cards one can also forecast correctly with probability $\frac{1}{2}$. In the latter case, flipping $n-1$ cards would lead to the same forecast as flipping $n$ cards.

The second observation is that the probability of a correct forecast does not grow with $n$ fast enough to justify flipping more cards. For example, consider an individual who has ten cards left to flip and contemplates whether to flip them all at once and make one forecast or flip five cards twice to make two forecasts. Recall that an individual's score is calculated as the number of correct forecasts less the number of incorrect forecasts. Therefore, the expected change in the score from flipping ten cards is $2p_{10}-1\approx 2\times 0.793 - 1 = 0.585$, whereas the expected change in the score from flipping five cards twice is $2(2p_5-1)\approx 2(2\times 0.707-1) = 0.829$. Similar analysis reveals that for any number of cards left to flip, maximum expected score is reached by splitting it into as many forecasts as possible.\footnote{To verify this claim, it is sufficient to consider all possible numbers of cards between 10 and 15, and all feasible splits of each of those numbers. Details are available from the authors upon request.} Thus, we obtain the following result.

\begin{proposition}
\label{prop_optimal_ind}
The unique optimal strategy for a risk-neutral participant under noncompetitive incentives is to flip five cards in each period.
\end{proposition} 

\begin{figure}[tbp]
{\includegraphics[width=3.0in]{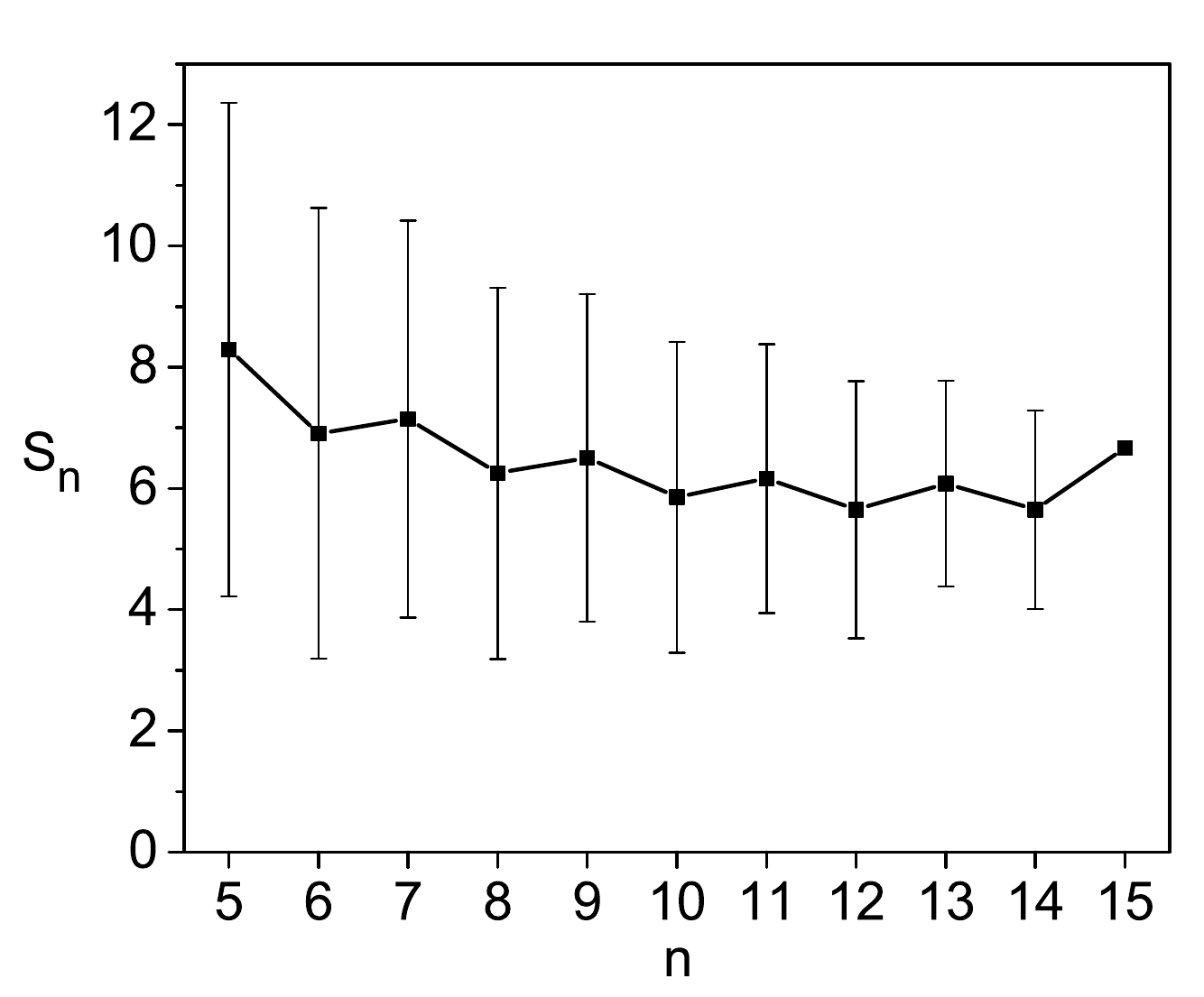}}
{\includegraphics[width=3.0in]{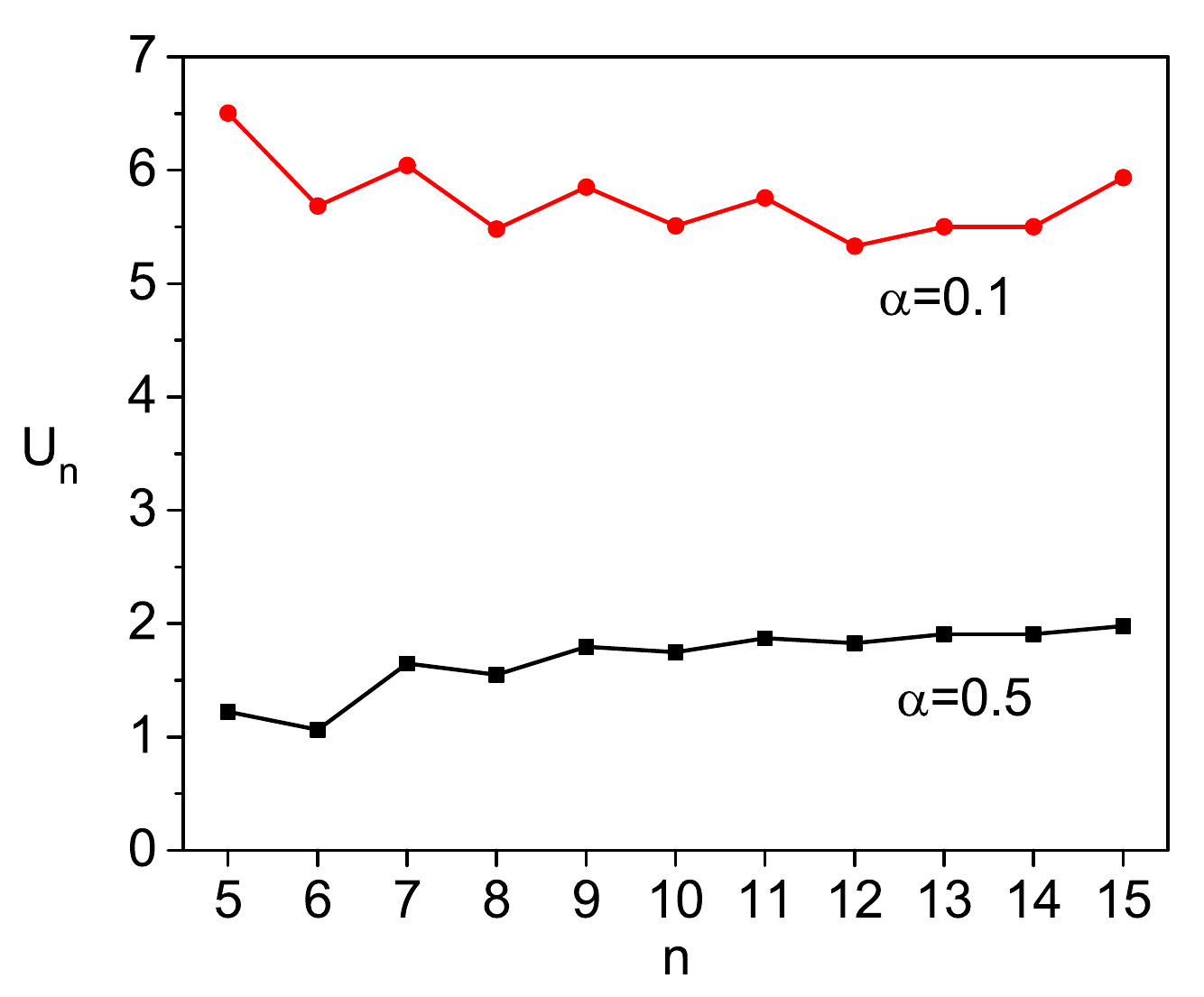}}
\caption{\textit{Left}: Expected score in a block, $S_n$, as a function of the number of cards flipped per period, $n$, calculated for the parameters of the experiment ($M=15$ and $N=100$) using Equations (\ref{pn_odd})-(\ref{ES}). The error bars show one standard deviation above and below the expected score for each $n$. \textit{Right}: Expected utility in a block, $U_n$, based on a CARA utility function $u(x)=\frac{1}{\alpha}(1-e^{-\alpha x})$, calculated using Equations (\ref{pn_odd}), (\ref{pn_even}), and (\ref{utility_RA}). The points are connected by straight lines for better visualization.}
\label{fig:ES_theory}
\end{figure}

For illustration, consider all stationary strategies such that a subject flips $n=5,\ldots,15$ cards per period. The subject's expected score after flipping a total of $N$ cards, ignoring the fact that $\frac{N}{n}$ may be non-integer, is
\begin{equation}
S_n=\frac{N}{n}(2p_n-1).  \label{ES}
\end{equation}
The left panel in Figure \ref{fig:ES_theory} shows the expected score for each value of $n$, calculated using the parameters of the experiment ($M=15$ and $N=100$). The error bars in the figure show one standard deviation above and below the expected score for each $n$. 

Consistent with Proposition \ref{prop_optimal_ind}, the expected score is highest when $n^*=5$---i.e. when the subject gathers the least amount of evidence---but its dependence on $n$ is rather flat, with expected scores between 6 and 8 for all allowed levels of card turning. The fluctuations in the dependence of $S_n$ on $n$ confirm that flipping an even number of cards $n$ is dominated by flipping $n-1$ cards. The variance in score decreases with $n$, confirming the trade-off between evidence gathering and returns in this environment. 

As a robustness check, consider also a risk-averse, expected utility maximizing subject with utility of money $u(\cdot)$. Restricting attention to stationary strategies, the corresponding optimal $n$ maximizes 
\begin{equation}
\label{utility_RA}
U_n = \sum_{k=0}^{\frac{N}{n}}\binom{\frac{N}{n}}{k}p_n^k(1-p_n)^{\frac{N}{n}-k}u\left(w\left(2k-\frac{N}{n}\right)\right).
\end{equation}
Here, $w$ is the piece rate (\$1.50 in the noncompetitive task), and $p_n$ is given by Equations (\ref{pn_odd}) and (\ref{pn_even}).\footnote{The goal of this limited analysis is to confirm relevant comparative statics with risk-averse subjects; a full analysis that allows for nonstationary strategies and history dependence is beyond the scope of the paper. Ultimately, how subjects actually behave in this setting is an empirical question. To understand equation (\ref{utility_RA}), consider a subject employing a stationary strategy $n$, i.e., flipping $n$ cards in each period. Ignoring the integer problem, such a subject will make $\frac{N}{n}$ forecasts, each of which can either be correct or incorrect. In (\ref{utility_RA}), we sum over all possible numbers of correct forecasts, $k\in\{0,1,\ldots,\frac{N}{n}\}$. The probability that $k$ forecasts are correct is given by the binomial formula, and the subject's payoff from $k$ correct forecasts is $w(2k-\frac{N}{n})$.} The right panel in Figure \ref{fig:ES_theory} shows expected utility as a function of $n$ using a constant absolute risk-aversion (CARA) utility of money, $u(x)=\frac{1}{\alpha}(1-e^{-\alpha x})$, for $\alpha=0.1$ and 0.5. The former value of $\alpha$ is close to the risk-neutral limit $(\alpha=0$), and expected utility is maximized at $n^*=5$ as in the case of risk neutrality. The latter value corresponds to a fairly risk-averse individual whose optimal strategy is $n^*=15$---the least risky one where the most evidence is gathered. In either case, the dependence of $U_n$ on $n$ is rather flat.\footnote{\label{fn_PT}A CRRA utility would be unsuitable for our purposes because it cannot accommodate negative payoffs. We also analyzed $U_n$ using, instead of the standard CARA utility function, a prospect theory-type value function, $u(x) = \max\{0,x\}^{\alpha}-\max\{0,-x\}^{\beta}$, with a reference point at zero and $0<\alpha<\beta$ corresponding to loss aversion. The results are qualitatively similar for a wide range of parameters.} 

We now turn to analyzing equilibrium behavior in the setting with a competitive rewards scheme. Given that $n^{\ast}=5$ maximizes a subject's expected score, it is a natural candidate for a symmetric Nash equilibrium under competitive rewards for risk-neutral players. However, since the strategy $n=5$ is also the riskiest (i.e., it leads to the highest variance in performance), it is not obvious that the choice of $n=5$ by a player is the best response to all other players choosing $n=5$. A safer strategy may decrease the player's chances of finishing first, but may simultaneously increase the player's chances of not being last. That is, the payoff from gathering more than the minimum evidence is not straightforward.

Due to intractability, we identify equilibria using numerical simulations. We focus on symmetric, stationary equilibria where all players choose to flip $n$ cards in each period within a block. Assuming all players except one flip some number of cards $n$ and the one indicative player flips $n'$ cards, we simulate a large number of realizations of scores of all players and calculate the indicative player's average payoff. We repeat this exercise for each $n'$ and find the indicative player's best response---the value $n'=n_{\rm opt}(n)$ that maximizes the average payoff for a given $n$. Going through all possible values of $n$, we look for symmetric equilibria satisfying $n_{\rm opt}(n^*)=n^*$. 

Using this approach, we find that $n_{\rm opt}(5)=5$. Deviations from $n=5$ when all other players flip 5 cards are not profitable---all subjects in the group choosing $n^*=5$ is a symmetric, stationary Nash equilibrium under risk-neutrality. Moreover, it is the only symmetric, stationary equilibrium since $n_{\rm opt}(n)=5$ for any $n\in\{5,\ldots,15\}$.

We also use similar simulations, with average payoffs replaced by average utility, to explore symmetric equilibria for risk-averse agents with the CARA utility function described above. The results change as expected. When the risk aversion parameter $\alpha$ is small, the symmetric, stationary equilibrium is still at $n^*=5$, but it switches to $n^*=15$---the safest strategy---when risk aversion becomes strong enough.\footnote{Similar changes happen, in a predictable way, when using the prospect theory-type value function described in footnote \ref{fn_PT}.}

\subsection{Conjectures}

Deciding how many cards to flip each period requires careful consideration. Although minimal evidence gathering is the optimal strategy for a risk-neutral subject, learning to play this strategy through standard reinforcement mechanisms is extremely difficult, if not impossible, especially without repetition. Fortunately, our research design does not rely on subjects' adopting theoretically optimal strategies; we can address our research question even if subject make non-optimal choices, including those consistent with probability matching. Subjects may also have different risk preferences. As a result, we expect substantial heterogeneity in behavior. Moreover, in our setting, subjects will likely view information about peers' strategies and outcomes as valuable and will respond to it.

Our experimental task has important features faced by investors and other risk-takers in the field. First, the relationship between strategies and outcomes, although it exists statistically, is weak and noisy. Second, outcomes are determined to a significant extent by luck, and disentangling luck and the effect of strategy is difficult.

The existing literature agrees that competitive incentives encourage more risk taking than comparable noncompetitive incentives, and we expect to find this result, on average, in our setting. 

\begin{conjecture}
\label{c1}
There is less evidence gathering under competitive incentives than noncompetitive incentives, in all feedback conditions.
\end{conjecture}

The literature provides little specific guidance, however, in terms of understanding the interaction between these incentives and the availability of information about peers' decisions and outcomes. Therefore, although we can lean on existing studies to build conjectures, our main research interest is the empirical relationship between incentives and feedback.

In the setting with information about peers' actions only, subjects cannot observe the effectiveness of others' strategies and, hence, we can assess whether subjects engage in ``pure'' imitation. Indeed, \cite{Karakostas-et-al:2021} find clustering of investment decisions in a risky gamble task when feedback about peers' previous investment decisions are disclosed. There are several possible reasons to expect such imitation. First, subjects are inexperienced in the environment, and making informed decisions is difficult. Moreover, due to reference group neglect \citep{Moore-Cain:2007}, subjects may believe that the task is difficult for them but not for others; hence, subjects may imitate their peers under the expectation that others know better. Second, it is plausible that subjects derive utility directly from conforming to the group norm or imitate to avoid anticipated social regret \citep{Cialdini-Goldstein:2004,Cooper-Rege:2011}. We expect any tendency towards the mean strategy to be especially pronounced in the setting with competitive rewards where relative positions matter and where, by conforming to their group's average behavior, subjects can reduce the riskiness of their overall strategy.\footnote{Under competitive rewards, it is also possible that subjects will engage in ``anti-imitation,'' whereby they attempt to break away from their peers by choosing a strategy that does not conform to the group mean. Such behavior is similar to choosing a riskier overall strategy and, therefore, it can be rationalized in a tournament setting \citep{Hvide:2002}.}

\begin{conjecture}
\label{c_strategies}
In treatments with information about peers' evidence gathering strategies only, there is imitation of peers' actions.
\end{conjecture}

In the setting with information about peers' scores only, there is limited scope for learning and informed strategy updating. Even though subjects do not observe how their peers achieved certain outcomes, they may make broad inferences about how to change their own strategies to improve their scores. For example, a subject pursuing a very safe strategy who learns that his or her score is substantially lower than the leader's score may adopt a riskier strategy in the next block. Conversely, an extreme risk-taker whose score turned out to be low may believe that he or she can mimic the group leader's success only with a safer strategy. Therefore, we expect subjects with scores further away from the group's maximum to change their strategy more dramatically. Subjects' relative positions are salient in the presence of score-only information, and we expect the effects of this information on strategy updating to be stronger in the setting with competitive rewards. This would be consistent with the results of \cite{Eriksen-Kvaloy:2017} who find that information on winners' earnings increases risk taking in tournaments. A similar ``keeping up with the winners'' finding is reported by \cite{Fafchamps-et-al:2015}, who find increased risk taking in the presence of information about peers' earnings in a noncompetitive risky choice task, although in that study subjects also had information on peers' investment decisions. It is also of interest to explore the behavior of the leader (i.e., whether the leader will choose to ``play it safe'' or adopt an even riskier strategy in the subsequent block).

\begin{conjecture}
\label{c_scores}
In treatments with information about peers' scores only, there is a tendency to change one's behavior more when one's outcome is further from the group maximum. The effect is stronger under competitive incentives than noncompetitive incentives.
\end{conjecture}

In the presence of information about peers' strategies and scores, subjects may imitate the leading scorer. \cite{Fafchamps-et-al:2015} find that individuals imitate the behavior of high-earning or ``winning'' group members. Even though the environment is noncompetitive, the salience of score rankings leads to implicit competition between individuals.\footnote{In a noncompetitive decision environment similar to the one used by \cite{Fafchamps-et-al:2015}, \cite{Celse-et-al:2021} find that competitive preference models, in which individuals gain utility by having investment or earnings above peers', best explain the experimental data.} The rationale is similar to imitation in the treatments with information about strategies only, except that the score leader's strategy (and not the average strategy) may be the attraction point. In the treatments with combined feedback about strategies and scores, subjects can observe patterns and, depending on their group, may observe different relationships between strategies and outcomes. Thus, learning will depend critically on what patterns arise. Again, the effect of this feedback on strategy updating is expected to be stronger in the setting with competitive rewards, when relative positions are particularly salient. 

\begin{conjecture}
\label{c_both}
In treatments with information about peers' evidence gathering and scores, there is a tendency to imitate the score leader and to change one's behavior following the observed pattern of dependence between evidence gathering and score. The effect is stronger under competitive incentives than noncompetitive incentives.
\end{conjecture}

\section{Results}
\label{sec:results}

In this section, we present and discuss our experimental findings. We first briefly discuss the frequency of flipping an even number of cards (a suboptimal strategy) and \textit{probability matching}, a suboptimal but pervasive guessing heuristic whereby subjects randomize their predictions to match the underlying probabilities of success. We offer a detailed analysis of probability matching in \ref{app_PM}. Importantly, we note that subjects' adoption of non-optimal strategies does not undermine our ability to address our main research question about the interaction between incentives and information about peers. Next, we describe the relationship between the uncertainty attitude measures and subjects' choices. We then describe the main results about the effects of the incentives and peer information on subjects' risk taking. Finally, we describe some patterns in the individual data that help to explain the observed treatment effects.

To assess the causal impact of information about peers, the analysis in this section focuses on subjects' evidence gathering decisions in block 2 where subjects are exposed to peer information for the first time. Because subjects are exposed to peer information again after blocks 2 and 3, the treatment effects in blocks 3 and 4 are confounded by the fact that subjects are responding to other subjects' responses to their earlier choices; hence, it is no longer possible to measure the causal effect of feedback in later blocks. Nevertheless, it may be of interest how subjects' behavior in each feedback and pay scheme condition evolves over time, and whether there are correlations between feedback content and decisions. These results are presented in Section \ref{sec:blocks34}.

\subsection{Flipping an even number of cards and probability matching}

To begin, we examine the basic properties of subjects' decisions. As discussed in the previous section, subjects should never choose to flip an even number of cards because it is dominated by flipping one fewer. In our experimental design, subjects may be forced to flip an even number of cards at the end of each block due to the restriction on flipping fewer than five cards in the final forecast. As such, we expect to see at most one even flip per subject per block. In fact, we observe that 33.3\% of all flips in the experiment are even---on average, subjects chose to make 4.14 even flips per block. The frequency of even flips declines over time, but only moderately, from 38.4\% in block 1 to 30.3\% in block 4.

The other, and perhaps more important, question is whether subjects base their forecasts on the majority color of the flipped cards. That is, do subjects make optimal predictions given their signals? Excluding cases in which an equal number of green and red cards are revealed, subjects forecast against the majority color in 18.5\% of their predictions. There is a very slight decline in the frequency of nonmajority guesses between the first and fourth block, from 19.0\% to 17.6\%, but the difference is not statistically significant. Only 8 of the 400 subjects (2\%) never made a nonmajority guess, and 41.3\% of subjects made 10 or more such guesses.

The presence of suboptimal, nonmajority guesses is a manifestation of the well-documented phenomenon of \textit{probability matching} that has been studied extensively in psychology, dating back to \cite{Estes:1950}. While observed prominently in our experiment, this behavior is not of central importance for our research questions, and a detailed analysis and discussion are presented in  \ref{app_PM}. The main conclusion from the analysis is that while probability matching reduces subjects' average payoffs compared to their payoffs had they always chosen optimally, this reduction is essentially a parallel downward shift that does not affect the risk--quantity trade-off of our forecasting task.

\subsection{Summary statistics}

\begin{table}[tbp]
\begin{center}
\caption{Summary statistics}
\label{tab:sumstat}
{\small
\begin{tabular}{lccccccccc}
\hline\hline
	& \multicolumn{4}{c}{\textbf{Noncompetitive Rewards}} & \multicolumn{4}{c}{\textbf{Competitive Rewards}}\\

\textbf{Peer Information:}	&	\textbf{None}	&	\textbf{Strategies}&	\textbf{Scores} 	&	\textbf{Both} &	\textbf{None}	&	\textbf{Strategies} 	&	\textbf{Scores} 	&	\textbf{Both}	\\
\hline
		&		&		&		&		&		&		&		&		\\
	& \multicolumn{8}{c}{\textit{Panel A: Block 1}}\\
\hline
\#\ of forecasts	&	13.11	&	13.31	&	12.58	&	12.79	&	13.40	&	13.20	&	13.30	&	13.09	\\
			&	(3.39)	&	(3.13)	&	(3.42)	&	(3.14)	&	(3.05)	&	(3.00)	&	(2.95)	&	(3.38)	\\
Score		&	6.22	&	5.14	&	5.18	&	6.09	&	6.56	&	5.70	&	7.25	&	6.09	\\
			&	(3.17)	&	(3.60)	&	(3.49)	&	(3.45)	&	(3.64)	&	(3.64)	&	(3.74)	&	(3.94)	\\
Luck in block 1 		&	-0.54		&	-1.67		&	-1.53	&	-0.65	&	-0.22	&	-1.02	&	0.46	&	-0.70	\\
		&	(3.08)	&	(3.85)	&	(3.52)	&	(3.41)	&	(3.64)	&	(3.71)	&	(3.53)	&	(3.77)	\\
			&		&		&		&		&		&		&		&		\\
	& \multicolumn{8}{c}{\textit{Panel B: Block 2}}\\
\hline
\#\  of forecasts	&	13.11	&	11.94	&	11.48	&	12.48	&	12.36	&	12.70	&	13.40	&	13.46	\\
			&	(3.85)	&	(3.42)	&	(3.73)	&	(3.32)	&	(3.32)	&	(3.55)	&	(3.48)	&	(4.08)	\\
Score		&	6.00	&	6.29	&	5.88	&	6.05	&	6.84	&	6.00	&	6.40	&	6.36	\\
			&	(4.00)	&	(3.14)	&	(2.65)	&	(3.72)	&	(3.03)	&	(3.53)	&	(2.86)	&	(3.80)	\\
			&		&		&		&		&		&		&		&		\\
	& \multicolumn{8}{c}{\textit{Panel C: Block 3}}\\
\hline
\#\ of forecasts	&	11.38	&	11.83	&	12.03	&	12.04	&	11.80	&	12.23	&	12.95	&	12.88	\\
&	(3.47)	&	(3.86)	&	(4.50)	&	(3.84)	&	(3.56)	&	(3.73)	&	(3.85)	&	(3.85)	\\
Score			&	5.56	&	6.86	&	6.23	&	6.39	&	6.42	&	7.28	&	6.10	&	6.30	\\
&	(2.76)	&	(2.56)	&	(2.86)	&	(2.73)	&	(3.20)	&	(2.88)	&	(3.22)	&	(2.87)	\\
&		&		&		&		&		&		&		&		\\
& \multicolumn{8}{c}{\textit{Panel D: Block 4}}\\
\hline
\#\  of forecasts	&	11.04	&	11.51	&	11.15	&	11.51	&	11.53	&	11.93	&	12.08	&	12.60	\\
&	(3.77)	&	(3.97)	&	(4.41)	&	(3.86)	&	(3.67)	&	(4.07)	&	(3.98)	&	(3.86)	\\
Score			&	5.98	&	6.54	&	6.80	&	6.09	&	6.56	&	6.68	&	6.28	&	6.28	\\
&	(2.46)	&	(2.68)	&	(2.71)	&	(2.53)	&	(2.78)	&	(3.22)	&	(2.60)	&	(2.83)	\\
&		&		&		&		&		&		&		&		\\
& \multicolumn{8}{c}{\textit{Panel E: Subject characteristics}}\\
\hline
Risk Aversion			&	8.71	&	7.66	&	7.75	&	8.83	&	8.56	&	8.73	&	8.95	&	8.80	\\
			&	(4.14)	&	(3.80)	&	(5.00)	&	(4.12)	&	(4.48)	&	(3.84)	&	(4.90)	&	(3.95)	\\
Ambiguity Aversion 			&	0.00	&	1.54	&	1.48	&	0.76		&	1.13	&	1.85	&	0.15	&	0.56	\\
			&	(4.49)	&	(3.63)	&	(5.30)	&	(4.80)	&	(5.75)	&	(5.00)	&	(5.78)	&	(5.74)	\\
Female		&	0.42	&	0.34	&	0.43	&	0.56	&	0.56	&	0.55	&	0.53	&	0.51	\\
			&	(0.50)	&	(0.48)	&	(0.50)	&	(0.50)	&	(0.50)	&	(0.50)	&	(0.51)	&	(0.50)	\\
\hline\hline
\end{tabular}
}
\captionsetup{font=small}
\caption*{Note: The first block of periods was identical across treatments. After the first block, all subjects received information about their own performance. We also implemented one of four peer information conditions: ``None'' indicates that subjects received no feedback about peers' evidence gathering strategies or scores; ``Strategies'' indicates that subjects received feedback about peers' evidence gathering strategies only; ``Scores'' indicates that subjects received feedback about peers' scores only; and ``Both'' indicates that subjects received feedback about peers' evidence gathering strategies and scores. Luck in block 1 is a subjects' actual score minus the expected score predicted by theory given the actual number of flipped cards. Standard deviations are reported in parentheses.}
\end{center}
\end{table}

Panels A through D of Table \ref{tab:sumstat} present summary statistics by treatment for blocks 1 through 4. Throughout our analysis, we use the number of forecasts in a block as a measure of subjects' evidence gathering behavior. More forecasts means that the subject collected less evidence by turning over fewer cards per period, whereas fewer forecasts means that the subject gathered more evidence by turning over more cards per period. Overall, an average subject turns over approximately 7 cards per period, leading to roughly 13 forecasts per block. The number of forecasts is not statistically different across treatments in block 1, a reassuring finding that confirms the experimental randomization.\footnote{A Kruskal-Wallis rank test yields a chi-squared statistic of 3.19 ($p=0.870$).} 

Panel A of Table \ref{tab:sumstat} also reports a measure of subjects' ``luck'' in block 1, defined as the difference between subjects' actual and expected scores. Expected score is calculated using the theoretical probability of a correct assessment, shown above in equation (\ref{ES}), given the actual number of cards flipped by subject. Subjects' luck in the first block did not vary statistically across treatments.\footnote{A Kruskal-Wallis rank test yields a chi-squared statistic of 8.43 ($p=0.297$).} Luck plays an important role in this noisy experimental environment and may influence subjects' ability to learn. A subject who was using an (\textit{ex ante}) suboptimal strategy and was lucky (or an \textit{ex ante} optimal strategy and was unlucky) will be less likely to update his or her strategy correctly, relative to a subject whose luck did not distort the signals about the theoretically superior strategies. 

Panel E of Table \ref{tab:sumstat} presents the summary statistics for the measures of risk and ambiguity aversion elicited at the start of the experiment. Risk aversion (RA) is measured as the number of safe choices in the risk aversion elicitation list. Ambiguity aversion (AA) is measured as the difference between the number of safe choices in the ambiguity aversion elicitation list and RA, where a positive difference suggests an aversion to ambiguity. Neither measure is statistically different across treatments, suggesting that subjects' attitudes towards uncertainty did not vary systematically by treatment.\footnote{Comparing the risk and ambiguity aversion measures across treatments, Kruskal-Wallis tests yield chi-squared statistics of 4.43 ($p=0.730$) and 5.97 ($p=0.544$), respectively.}

Ideally, the risk and ambiguity aversion measures would capture the underlying preferences of individual subjects and, as a result, explain subjects' decisions in the first block of the experiment. To test this hypothesis, we regressed the number of forecasts in the first block on the risk and ambiguity aversion measures. As expected, the coefficient estimates on both measures are negative, although only the coefficient for risk aversion is marginally statistically significant ($p=0.068$), and the $R^2$ of the regression is 0.010. We also checked whether evidence gathering behavior in the first block is affected by gender. Numerous studies have documented differences in the risk attitudes of men and women \citep[for a review, see, e.g.,][]{Croson-Gneezy:2009}. However, we do not find a correlation between risk-reducing evidence gathering and gender, whether risk and ambiguity aversion are controlled for or not; the coefficient estimate on an indicator for female is $-0.14$ ($p=0.653$) and $-0.10$ ($p=0.756$), respectively. 

Although subjects' risk aversion measure is correlated with their evidence gathering behavior in the first block, very little of the variation in subjects' strategies is explained by individuals' observable characteristics. Therefore, in the analysis that follows, we use the subjects' decisions in the first block (i.e. the number of forecasts in block 1) as a subject-specific control.

Recall that block 1 is identical across treatments; therefore, the content of the feedback in block 2 is orthogonal to the treatments. In contrast, in blocks 3 and 4, the feedback is no longer orthogonal to treatments because subjects are now responding to what happened in the previous block where their behavior (and hence, the resulting feedback) is itself correlated with the treatment. For this reason, to assess most cleanly the causal impact of the information treatments and feedback content, we focus on the analysis of evidence gathering in block 2 where subjects were exposed to information about their peers for the first time. Behavior in blocks 3 and 4 is still of interest as it allows us to see where strategies in the corresponding treatment converge over time. We present the analysis of behavior in blocks 3 and 4 in Section \ref{sec:blocks34}.

Across all of the treatments, we find that there is a statistically significant increase in the average level of evidence gathering between the first and second block; this increase can be primarily attributed to subjects who gathered less than the average amount of evidence in the first block.\footnote{In a regression of the change in the number of forecasts between blocks 1 and 2 on the number of forecasts in block 1 and a constant, with standard errors clustered by group, we estimate a positive constant and a negative coefficient on the number of forecasts in block 1 ($p<0.001$ for both).}  

\subsection{Main results}
\label{sec:main_results}

\begin{table}[tbp]
\begin{center}
\caption{Evidence gathering, incentives, and peer information, block 2}
{\small
\begin{tabular}{lccccc}
\hline\hline
\multicolumn{4}{l}{Dependent variable: \textit{\#\ of forecasts in block 2}} & &\\
	&	\ref{tab:baseline}.A	&	\ref{tab:baseline}.B	&	\ref{tab:baseline}.C	&	\ref{tab:baseline}.D	&	\ref{tab:baseline}.E	\\
\hline
1(Competitive)	&	0.89*	&	0.65*	&	0.51	&	-1.15*	&	-1.14*	\\
	&	(0.49)	&	(0.37)	&	(0.34)	&	(0.68)	&	(0.68)	\\
1(Peer information)	&		&		&		&	-0.83	&		\\
	&		&		&		&	(0.58)	&		\\
1(Competitive)$\times$1(Peer information)	&		&		&		&	2.13***	&		\\
	&		&		&		&	(0.78)	&		\\
1(Peer information: strategies only)	&		&		&		&		&	-1.28**	\\
	&		&		&		&		&	(0.63)	\\
1(Peer information: scores only)	&		&		&		&		&	-1.20*	\\
	&		&		&		&		&	(0.63)	\\
1(Peer information: strategies 	&		&		&		&		&	-0.43	\\
\hspace*{2em}\& scores)	&		&		&		&		&	(0.66)		\\
1(Competitive)$\times$1(Peer &		&		&		&		&	1.94**		\\
\hspace*{2em}information: strategies only)	&		&		&		&		&	(0.81)		\\
1(Competitive)$\times$1(Peer	&		&		&		&		&	2.29**		\\
\hspace*{2em}information: scores only)	&		&		&		&		&	(0.96)		\\
1(Competitive)$\times$1(Peer	&		&		&		&		&	2.16**		\\
\hspace*{2em}information: strategies \& scores)	&		&		&		&		&	(0.91)		\\
\#\ of forecasts in block 1	&		&	0.80***	&	0.79***	&	0.79***	&	0.80***	\\
	&		&	(0.06)	&	(0.06)	&	(0.06)	&	(0.06)	\\
Luck in block 1	&		&		&	0.28***	&	0.28***	&	0.27***	\\
	&		&		&	(0.05)	&	(0.05)	&	(0.05)	\\
Constant	&	12.28***	&	1.94**	&	2.33***	&	2.95***	&	2.90***		\\
	&	(0.35)	&	(0.84)	&	(0.77)	&	(0.90)	&	(0.90)		\\
\hline
$N$			&	400	&	400	&	400	&	400	&	400	\\
Pseudo $R^2$	&	0.01	&	0.09	&	0.11	&	0.11	&	0.12	\\
\hline\hline
\end{tabular}
\label{tab:baseline}
}
\captionsetup{font=small}
\caption*{Note: Tobit regression results using data from all treatments. Standard errors, clustered by group, are reported in parentheses; there are 152 clusters in each regression. Group size is five in all treatments except the treatments without feedback, where the group size is one because subjects were not exposed to information about their peers. *, ** and *** indicate statistical significance at the 10\%, 5\% and 1\% levels, respectively.}
\end{center}
\end{table}

With eight experimental conditions across two blocks, it is difficult to glean a simple conclusion about the effect of information about peers' strategies and outcomes on evidence gathering from only the summary statistics in Table \ref{tab:sumstat}. Examining the average number of forecasting attempts in block 2, it appears that competitive rewards are associated with less evidence gathering when subjects receive information about their peers. Comparing the average number of forecasts under noncompetitive and competitive incentives for each feedback condition, we find that the difference is only significant with feedback about scores only ($p=0.048$; here and below, for all comparisons we use the two-sided Wald test with standard errors clustered by group). Comparing strategies across information conditions within each incentive scheme, we find significantly less evidence gathering with feedback about peers' scores as compared to no feedback, under noncompetitive rewards ($p=0.023$). Of course, it is critical to recognize that Table \ref{tab:sumstat} summarizes only the \textit{average} strategies and scores of all subjects within a treatment. Regression analysis allows us to identify treatment effects in the data while accounting for subject-level heterogeneity.

Table \ref{tab:baseline} presents baseline results comparing subjects' evidence gathering under competitive and noncompetitive rewards with and without information about peers' strategies and scores. The regressions build up from a simple specification to one that captures the full experimental design. In all regressions, the dependent variable is the number of forecasting attempts by individual subjects in the second block. Each column reports results of a Tobit specification that accounts for the fact that forecasts are bounded between 7 and 20, with significant bunching at the boundaries; in our experiment, the number of forecasting attempts is 7 in 5.75\% of observations and 20 in 8.75\% of observations. Standard errors are clustered by group.\footnote{In treatments with feedback, groups are defined as the five subjects who were presented with information about each others' scores, strategies or both; in treatments without feedback about peers, groups include only a single subject, since these individuals are not exposed to information about other participants.} 

We begin by examining the common notion that competitive reward schemes lead to more risk taking that would appear, in our context, as less risk-reducing evidence gathering. The results reported in column \ref{tab:baseline}.A provide this comparison pooling all feedback conditions together. The coefficient estimate on the indicator for competitive incentives is positive and marginally statistically significant ($p=0.068$). In terms of magnitude, competitive incentives are associated with 0.89 more attempts in block 2, on average, which is nontrivial. Recall that one attempt can increase or reduce a subject's score by one. The average score in block 2 was 6.23 (cf. panel B of Table \ref{tab:sumstat}); therefore, a difference in 0.89 attempts translates into potential changes in score of about 14\%. 

Drawing from our earlier discussion of subjects' strategies and the extant literature, we expect underlying attitudes towards risk taking, as well as experiences in the first block, to vary across subjects. In column \ref{tab:baseline}.B, we include the number of forecasts in block 1 as a control for each subjects' risk taking in the first block. The estimated coefficient on the indicator for competitive incentives is still positive, but slightly smaller than in column \ref{tab:baseline}.A and also marginally statistically significant ($p=0.081$). The coefficient estimate for the risk taking from the first block is positive and statistically significant ($p<0.001$), suggesting it captures subject-level differences well.

One might also ask whether subjects' own experiences in the first block influence their choices in the second block. To that end, in column \ref{tab:baseline}.C, we add a measure of the subject's ``luck'' in the first block, defined as the difference between the subject's actual score and the expected score predicted by theory, given the number of cards flipped. The coefficient estimate for the competitive incentives indicator remains positive, but it is no longer statistically significant at conventional levels. The luck measure itself has explanatory power---the coefficient is positive and statistically significant ($p<0.001$)---suggesting that subjects decrease their evidence gathering in response to previous good luck.

Whereas columns \ref{tab:baseline}.A, \ref{tab:baseline}.B and \ref{tab:baseline}.C address the question of the average impact of competitive rewards, our main research question focuses on the interaction of incentives and peer information. We first analyze the effect of peer information in general, without separating different types of feedback. In column \ref{tab:baseline}.D, we report results from a regression that includes an indicator for the presence of any peer information (scores, strategies or both) and its interaction with the indicator for competitive incentives. When no peer information is available to subjects, competitive rewards are associated with more evidence gathering than noncompetitive rewards; the coefficient on 1(Competitive) is negative, relatively large ($-1.15$) and marginally statistical significant ($p=0.092$). In contrast, when peer information is available, competitive rewards are associated with less evidence gathering; the sum of the coefficients on 1(Competitive) and 1(Competitive)$\times$1(Peer Information) is positive ($0.98$) and statistically significant ($p=0.011$).

We can also hold the incentives fixed and isolate the effect of feedback. Peer information appears to have no significant impact on evidence gathering in the noncompetitive setting; the coefficient on 1(Peer information) is negative and not statistically significant ($p=0.150$). At the same time, peer information leads to less evidence gathering when subjects face competitive rewards; the sum of the coefficients on 1(Peer information) and 1(Competitive)$\times$1(Peer Information) is positive (1.30), and statistically significant ($p=0.013$).

Result \ref{res1} summarizes our findings from the preliminary analysis presented in columns \ref{tab:baseline}.A--D:

\begin{result}
\label{res1}
(a) There is no evidence of a robust effect of competitive incentives on the average level of evidence gathering. 

(b) In the absence of any information about peers, there is more evidence gathering under competitive incentives than under noncompetitive incentives, but the difference is only marginally statistically significant.

(c) In the presence of information about peers, there is less evidence gathering under competitive incentives than under noncompetitive incentives.
\end{result}

Results \ref{res1} describes a surprising finding vis-\`{a}-vis Conjecture \ref{c1}. In contrast to the conventional view that tournament-style incentives schemes are always associated with more risk taking relative to noncompetitive incentives, we find that competitive rewards may lead to behavior associated with \textit{less} risk taking when subjects receive no information about other competitors. This result is overturned in the presence of peer information, when competitive rewards are associated with less risk-reducing evidence gathering. In short, the availability of information about peers appears to matter critically. Of course, the coarse indicator for peer information in column \ref{tab:baseline}.D may obscure the heterogeneous impact of different types of feedback, which we consider next.

To explore the impact of different types of peer information on evidence gathering, we report the results of a regression with indicators for feedback about peers' strategies, scores or both under each incentive scheme in column \ref{tab:baseline}.E. To begin, we compare the effect of competitive and noncompetitive incentives while holding fixed the information available to subjects. In the absence of peer information, competitive compensation is associated with more evidence gathering than a noncompetitive pay scheme, with marginal statistical significance ($p=0.092$). When information about peers' strategies, scores or both is available, subjects engage in less evidence gathering when they face competitive rewards; for each type of information, the sum of the coefficient estimates on the indicator for competitive rewards and its interaction with the indicator for specific information type is positive and marginally significant ($p=0.075$ for strategies only, $p=0.094$ for scores only, and $p=0.091$ for both strategies and scores). Different types of feedback appear to have the same effect on evidence gathering; there are no statistically significant differences between the effects of information about strategies, scores or both. 

We can also estimate the impact of each type of peer information while holding fixed the incentive scheme. Under noncompetitive rewards, subjects engage in more evidence gathering when they receive feedback about their peers' strategies only ($p=0.041$) or, with marginal significance, scores only ($p=0.059$), but not when they receive information about both strategies and scores ($p=0.518$). In contrast, under competitive rewards, the combined information on peers' strategies and scores is associated with significantly less evidence gathering ($p=0.005$), whereas the effects of feedback on peers' strategies only or scores only are not statistically significant ($p=0.207$ and $p=0.127$, respectively). We summarize the findings from column \ref{tab:baseline}.E in Result \ref{res2}:

\begin{result}
\label{res2} 
(a) Comparing competitive and noncompetitive incentives, competitive rewards are associated with less evidence gathering for each type of peer information; the effects are marginally significant. 

(b) When incentives are noncompetitive, information about either peers' strategies or peers' scores is associated with more evidence gathering; however, the availability of information about both strategies and scores has no effect. 

(c) When incentives are competitive, information about either peers' strategies or peers' scores does not affect evidence gathering; however, information about both strategies and scores is associated with less evidence gathering. 
\end{result}

Result \ref{res2}(a) decomposes Result \ref{res1}(b): although the differences are noisy, competitive rewards lead to less evidence gathering than the noncompetitive scheme for each type of peer information. 

Taken together, our results so far suggest that tournament-style compensation does not inherently induce more risk taking in the form of less evidence gathering. In the presence of information about peers' strategies, scores or both, subjects facing relative performance rewards may engage in less costly evidence gathering than they would with noncompetitive rewards. However, in settings without peer information, competitive rewards may be associated with \textit{more} evidence gathering. Overall, the presence or absence of information matters critically in terms of how competitive rewards affect risk taking. That is, to achieve a particular organizational objective, one needs to consider both the incentives and the availability of information to competitors. Moreover, the nature of the feedback matters---holding incentives fixed, the type of peer information may affect evidence gathering.

It is instructive to relate our findings to those in the existing literature. Whereas prior studies typically focus on one type of feedback and/or one type of incentives, our treatments provide a systematic exploration of the interaction of different types of feedback with both types of incentives. 

Under competitive incentives, \cite{Eriksen-Kvaloy:2014} and \cite{Eriksen-Kvaloy:2017} found an increase in risk taking due to more frequent information about peers' strategies and outcomes and in the presence of information about the winner's strategies, respectively. Both studies used peer information consistent with our combined strategies and scores information condition, and like those studies, we too find a positive effect of feedback on risk taking. However, we also explore each type of peer information in isolation and find that neither source of peer information has an effect on risk taking in isolation. Thus, our findings provide some degree of robustness to variation in the task \citep[both prior studies used a variant of the investment task of][]{Gneezy-Potters:1997}, but also show that in order to have an effect the available peer information must be sufficiently rich.

Under noncompetitive incentives, \cite{Celse-et-al:2021} found a positive effect of feedback about a peer's investment decision and payoff on risk taking. In contrast, we do not find such an effect in our experiment and, moreover, we find a negative effect on risk taking when examining each type of peer information in isolation. We also do not observe a ``keeping-up-with-the-winners'' effect found by \cite{Fafchamps-et-al:2015}. However, these authors also find a negative effect of peers' lottery outcomes, which is consistent with our result on the negative effect of feedback in the scores-only treatment. \cite{Karakostas-et-al:2021} vary feedback content by providing an initial anchor---investment decisions from a previous study---that can be high or low and also vary whether or not social information is provided within the experiment (i.e., whether or not peers' investment is disclosed after each round). They find more risk taking in the high anchor treatment, with and without peer investment feedback. This is related---although not entirely parallel---to our strategies-only condition where, in contrast, we find an overall negative effect on risk taking.

Our results also raise questions about \textit{how} feedback reverses the unexpected difference between subjects' risk-related choices in competitive and noncompetitive settings. For example, does information about peers' scores motivate more aggressive decision-making by giving average competitors a sense of how far they lag behind the leaders? Does information on risk taking give subjects a view into possible strategies to imitate? Does information about strategies and scores allow competitors to identify a path to success? For each type of peer feedback, there may be different mechanisms driving the higher risk taking that we observe under competitive compensation schemes. In the following sections, we shed light on possible mechanisms.

As robustness checks, we ran three alternative specifications of the regressions in Table \ref{tab:baseline}; detailed results are available from the authors upon request. First, we added subject-level controls for gender and the risk-aversion and ambiguity-aversion measures. Given that the number of attempts in block 1 already subsumes individual, time-constant factors, these controls likely add little to our analysis and, indeed, we find coefficient estimates that are similar in magnitude and statistical significance to those reported here. Second, we re-ran the regressions in Table \ref{tab:baseline} using a pooled OLS instead of a pooled Tobit specification. We believe the Tobit specification is justified since 5.75\% of observations are left-censored (23 out of 400 made 7 attempts in block 2) and 8.75\% are right-censored (35 out of 400 made 20 attempts). Yet, our main results in Table \ref{tab:baseline} practically do not change if we use pooled OLS instead. Finally, we re-ran the regressions in Table \ref{tab:baseline} using a difference-in-differences specification where the dependent variable is the difference in the subject's number of attempts between blocks 2 and 1, and we do not control for the number of attempts in block 1. It is worth noting that the difference-in-differences specification is more restrictive than (the OLS version of) the one we report because it imposes a coefficient of 1 on the number of attempts in block 1, whereas our preferred specification leaves the coefficient free. In this case, without treatment controls (columns A, B, C), the results are directionally the same as in Table \ref{tab:baseline}, but coefficient estimates are no longer statistical significant at conventional levels. However, when we add treatment controls (columns D and E), the results become very similar to those in Table \ref{tab:baseline}. 

\subsection{Information about peers' scores}
\label{sec:scoresonly}

\begin{table}[tbp]
\begin{center}
\caption{Distance to the best- and worst-scoring peers, block 2}
{\small
\begin{tabular}{lccc}
\hline\hline
\multicolumn{3}{l}{Dependent variable: \textit{\#\ of forecasts in block 2}} \\
          &       \ref{tab:scores}.A    &         \ref{tab:scores}.B    &         \ref{tab:scores}.C   \\
\hline
1(Competitive)          &       -0.79    &       0.87    &   1.45*   \\
                    &      (0.55)    &     (0.56)    &     (0.79)   \\
Distance to the group's highest 		&        0.02    &                &         \\
   \hspace*{2em}  score in block 1                 &      (0.16)    &                &        \\
1(Competitive)$\times$Distance to the &        0.26*** &                &       \\
 \hspace*{2em} group's highest score in block 1         &      (0.09)    &                &        \\
Distance to the group's lowest		&                 &       0.10    &          \\
    \hspace*{2em}  score in block 1                    &                 &     (0.13)    &        \\
1(Competitive)$\times$Distance to the &                 &      -0.01    &         \\
  \hspace*{2em} group's lowest score in block 1         &                 &     (0.16)    &        \\
RelDist in block 1 &                 &          &  0.56       \\
           &                 &         &  (1.00)      \\
1(Competitive)$\times$RelDist in block 1 &                 &          &  -1.59*       \\
           &                 &         &  (0.85)      \\
\#\ of forecasts in block 1      &        0.88*** &       0.82*** &       0.85***\\
                    &      (0.09)    &     (0.11)    &     (0.11)   \\
Luck in block 1				&        0.56*** &       0.38**  &       0.46** \\
                    &      (0.19)    &     (0.14)    &     (0.12)   \\
Constant            &        1.07    &       1.18    &       1.20   \\
                    &      (1.30)    &     (1.26)    &     (1.19)   \\
\hline
$N$                   &           80    &          80    &          80   \\
Pseudo $R^2$                &        0.24    &       0.23    &       0.23   \\
\hline\hline
\end{tabular}
\captionsetup{font=small}
\caption*{Note: Tobit regression results using data from treatments in which subjects received information about peers' scores only. Group-level clustered standard errors are reported in parentheses; there are 16 clusters in each regression. *,  ** and *** indicate statistical significance at the 10\%, 5\% and 1\% levels, respectively.}
\label{tab:scores}
}
\end{center}
\end{table}

We start with an analysis of an environment in which subjects learn about peers' outcomes, but not the strategies that underlie those successes or failures. Outside of the laboratory, this setting aligns with situations in which, for example, colleagues share stories about their portfolios' returns, the volume of completed deals, or their year-end totals, without describing \textit{how} they achieved these successes. Alternatively, workers may describe their losses or shortfalls without explaining what led to these failings.

Table \ref{tab:scores} reports regression results using the number of forecasts in block 2 as the dependent variable and examines the two treatments---with noncompetitive and competitive rewards---in which subjects received only score-related feedback about their peers. As in Table \ref{tab:baseline}, a Tobit specification accounts for the fact that forecasts are bounded between 7 and 20. All regressions include subject-specific controls for the number of forecasts and luck in block 1. To understand how feedback about a subject's relative position in the group affects evidence gathering, in column \ref{tab:scores}.A, we include a measure of the distance between a subject's score and the highest score in the group and its interaction with the indicator for competitive incentives. Similarly, in column \ref{tab:scores}.B, we include the distance to the lowest score in the group. Finally, in column \ref{tab:scores}.C, we include a relative distance variable defined as ${\rm RelDist} = \frac{{\rm OwnScore}-{\rm MinScore}}{{\rm MaxScore}-{\rm MinScore}}$, where ${\rm OwnScore}$, ${\rm MinScore}$ and ${\rm MaxScore}$ are, respectively, the subject's own score, minimum and maximum group score in block 1.\footnote{We define RelDist=0.5 if all subjects in the group have the same score. This never occurs in our data.} A larger relative distance indicates a higher performance relative to one's peers, with RelDist$=0$ (1) corresponding to the lowest (highest) score in the group.

In settings with noncompetitive rewards, the distance between group members' scores appears to have little influence on decisions; across the three columns in Table \ref{tab:scores}, the coefficient estimates on the (uninteracted) distance measures are not statistically significant. The interaction of distance to the maximum score with the competitive indicator is positive and significant in \ref{tab:scores}.A ($p=0.003$), implying that subjects' reaction to this distance is stronger under competition. At the same time, the overall effect of the distance to the maximum under competition is quite noisy ($p=0.118$ for the sum of coefficient estimates on the distance to the leader and its interaction with competitive rewards). 

There are no similar effects for the distance from the lowest score in column \ref{tab:scores}.B. In column \ref{tab:scores}.C, the marginally significant positive coefficient on competitive rewards ($p=0.071$) suggests that lowest-scoring subjects (i.e., those with ${\rm RelDist}=0$) take more risk under competitive incentives. However, the negative coefficient on the interaction of RelDist with competitive rewards shows that this effect disappears ($p=0.417$ for the sum of coefficient estimates on RelDist and its interaction with competitive rewards) as subjects move away from the lowest score (and closer to the highest score) consistent with columns A and B.\footnote{As a falsification exercise, we estimate the same specification as the one reported in column \ref{tab:scores}.C examining only treatments in which subjects received \textit{no} information about their peers. Using these data, none of the distance measures or interactions is statistically significant.}%$p=0.116$ 

We also examine responses to the open-ended questions described in Section \ref{sec:design}. Subjects in the treatment with competitive rewards and information about peers' scores are more likely to claim that they followed a strategy, relative to those in the noncompetitive treatment with similar information (question (i)). Approximately 73\% of the subjects claim that their strategy changed over the rounds of the experiment (question (ii)). Responses suggest that peers influence those changes; 23\% of the subjects claim that they changed strategies after observing peers' performance, with 4\% (11\%) stating that they switched to a safer (riskier) strategy (question (iii)).\footnote{Note that 4\% and 11\% do not add up to 23\% because the remaining subjects did not unambiguously state how their strategy changed.} As expected for treatments involving only performance information, no subject states that they were imitating their peers (question (iv)). 

We summarize the findings about subjects' relative position, based on the estimates in Table \ref{tab:scores}, as follows:

\begin{result}
\label{res3} Although the results are noisy, there is evidence that a subject's relative position---distance to the highest score---is associated with less evidence gathering under competitive incentives. 
\end{result}

Result \ref{res3} suggests that subjects adopt riskier strategies in competition, as compared to a setting with noncompetitive rewards, when they need to overcome a larger gap to catch up with the leader. This ``Hail Mary'' behavior is consistent with that found by \cite{Eriksen-Kvaloy:2017} and \cite{Kirchler-et-al:2018}.

As a robustness check, we also separately examined how the behavior of the highest and lowest-scoring subjects changes between blocks 1 and 2. For this analysis, we use data from the treatments without peer information and with information about peers' scores only, under both types of incentives. We regress the difference between a subject's number of forecasts in the second and first block on the indicators for peer information and competitive rewards and their interactions, along with the controls for the first block forecasts and luck.\footnote{For brevity, we do not report the full results here; they are available from the authors by request.} We identify how a change in a subject's evidence gathering between blocks 1 and 2 is affected by \textit{learning} that he or she had either the highest or lowest score in the group in block 1. Subjects who learn that they are leading do not adjust their behavior across the first two blocks. Under noncompetitive incentives, subjects who learn that their score was the lowest gather more evidence between blocks 1 and 2, although the effect is only marginally significant ($p=0.092$). The effect is reversed for lowest-scoring subjects facing competitive incentives, but it is not statistically significant ($p=0.201$).

\subsection{Information about peers' strategies}

Information about peers' actions might lead to the convergence of subjects' strategies within groups. To consider this possibility, we calculate the difference between the standard deviations of submitted forecasts in the first and second block for each group. Overall, we find no evidence that the presence of feedback about evidence gathering led subjects towards a common strategy. 

More specifically, we first hold fixed the type of information and compare the change in the dispersion of forecasting strategies under the two incentive structures. In the treatments without peer information, the group-level standard deviation declines in the noncompetitive treatment and is unchanged in the competitive setting; however, the difference between these changes is not significantly different from zero $(p=0.273)$. When subjects receive information about peers' evidence gathering, the difference is even smaller in magnitude (both changes are negative) with an even larger standard error.

The changes are similarly small when we hold fixed the incentive structure and consider the impact of peer information. Under both competitive and noncompetitive schemes, comparing the treatments without peer information and with only information about peers' strategies, the differences in standard deviations in the first and second blocks are small and not statistically different from zero ($p=0.731$ for noncompetitive rewards, $p=0.174$ for competitive rewards). 

%A comparison of treatments without peer information and with only information on risk taking yields $p=0.731$ when subjects face noncompetitive rewards and $p=0.174$ when subjects face competitive rewards. 

% If you don't see convergence, could be other stuff going on. we look for some of those things.
% does tournament create incentives for anti-imitation. might be strategic. but we also don't see it in piece-rate.
% could be noisy. but we see strong effects with scores, so people do seem to react to some stuff. 
% feedback on scores has immediate impact on payoff relevant things. feedback on strategies is less direct, less obvious pay-off relevance. 

%looking at variance is analogous to Cooper/Rege looking at the role of the majority. moving towards the median guy is like following the majority 

One might ask whether subjects who learn about their peers' choices then attempt to imitate some of those strategies. To consider this question, we examine the impact of the mean, median, minimum and maximum number of forecasts by peers and of the difference between those measures and a subject's number of forecasts in block 1, using a specification similar to that reported in column \ref{tab:scores}.A for scores; however, we find no statistically significant relationships between subjects' own forecasts and those of their peers.

Overall, we find little evidence for Conjecture \ref{c_strategies} that the presence of information about peers' evidence gathering influences subjects' choices. In the text responses to the follow-up questions, 68\% of the subjects claim that their strategy changed over the rounds of the experiment (question (ii)). Only 16\% of the subjects write that these changes occurred after observing peers' strategies, with 11\% (4\%) of subjects stating that they switched to a safer (riskier) strategy (question (iii)). Moreover, when asked directly, only 1\% of the subjects indicate that they imitated other players' strategies (question (iv)). This pattern of responses did not vary between the competitive and noncompetitive treatments. 

% speculate about why no imitation. treatments 4 and 14 provide more information. environment is complex.
% look at evidence in other papers for imitation (cooper & rege; groups of 6; feedback about each player). utility from doing what others are doing. don't see that here. look at other papers about imitation.... 
% two players vs 5 players. less obvious who to imitate (if anyone). more complicated than with two players.
% social influence much less obvious. if you get utility from doing what other people do, if those other people are doing different things, who do you follow. 

% we looked at following the highest risk guy and the lowest risk guy and the mean risk guy. but didn't find anything. this could mean that people aren't following. or it could mean that people are very heterogeneous in who they follow, so we aren't picking it up in these data. 

% ambiguous environment. complex. 

We summarize these observations about the availability of feedback about peers' strategies as follows:

\begin{result}
\label{res5}
In both the competitive and noncompetitive settings, information about peers' evidence gathering strategies alone does not lead to convergence of subjects' strategies, and we find no evidence that subjects imitate their peers' strategies.
\end{result}

\subsection{Information about peers' strategies and scores}
\label{sec:strategies_scores}

\begin{table}[tbp]
\begin{center}
\caption{Strategies and scores, block 2}
{\small
\begin{tabular}{lccccc}
\hline\hline
\multicolumn{3}{l}{Dependent variable: \textit{\#\ of forecasts in block 2}} &       &     \\
      &        \ref{tab:corr}.A     &        \ref{tab:corr}.B     &        \ref{tab:corr}.C    &         \ref{tab:corr}.D   &          \ref{tab:corr}.E   \\
\hline
1(Competitive)            					&      1.04     &      0.39     &      1.33**  &      0.95   &        1.13** \\
                      						&    (0.63)     &    (1.04)     &    (0.57)    &     (1.09)   &      (0.51)   \\
RelDist in block 1 		  				&                &      -2.24**    &               &       -0.99   &                \\
           			&                &    (1.12)     &               &     (1.23)   &                \\
1(Competitive)$\times$RelDist in block 1 		&                &      0.77     &               &       0.77   &                \\
         &                &    (1.48)     &               &     (1.45)   &                \\
Correlation of forecasts and	         			&                &                &     2.36*** &      2.17** &       2.11***\\
       \hspace*{2em}   scores in block 1                  	&                &                &    (0.74)    &     (0.87)   &      (0.76)   \\
1(Competitive)$\times$Correlation of forecasts     	&                &                &      -2.02    &      -1.85   &        -2.94** \\
      \hspace*{2em}  and scores in block 1                    &                &                &    (1.22)    &     (1.43)   &      (1.28)   \\
\#\ of forecasts by subject with highest	  		&                &                &               &               &        0.264*  \\
     \hspace*{2em}  score in block 1                       	&                &                &               &               &      (0.14)   \\
\#\ of forecasts by subject with lowest	   		&                &                &               &               &        0.13   \\
     \hspace*{2em}  score in block 1                        	&                &                &               &               &      (0.10)   \\
\#\ of forecasts in block 1        				&      0.77***  &      0.79***  &      0.74*** &       0.74***&        0.68***\\
                  							&    (0.11)     &    (0.11)     &    (0.10)    &     (0.10)   &      (0.10)   \\
Luck in block 1  						&      0.24***  &      0.59***  &      0.18**  &       0.23  &        0.17** \\
                     							&    (0.08)     &    (0.17)     &    (0.08)    &     (0.15)   &      (0.08)   \\
Constant            						&      2.83**   &      2.28*    &      2.75**  &       3.26** &       -1.80   \\
                    							&    (1.35)     &    (1.25)     &    (1.18)    &     (1.30)   &      (1.65)   \\
\hline
$N$                     &        155     &        155     &        155    &         155   &          155   \\
Pseudo $R^2$          &      0.098     &      0.102     &      0.11    &       0.111   &        0.119  \\ 
\hline\hline
\end{tabular}
\captionsetup{font=small}
\caption*{Note: Tobit regression results using data from treatments in which subjects received information about peers' strategies and scores. Group-level clustered standard errors are reported in parentheses; there are 31 clusters in each regression. *, ** and *** indicate statistical significance at the 10\%, 5\% and 1\% levels, respectively.}
\label{tab:corr}
}
\end{center}
\end{table}

Subjects who are presented with information about both the strategies and scores of their peers in block 1 receive the richest feedback in our experimental design. Given this information, subjects can evaluate their own success and choices relative to other subjects in their group. Moreover, these subjects can assess the \textit{overall} relationship between costly evidence gathering and overall success. 

Table \ref{tab:corr} presents regression results with our evidence gathering measure,  the number of forecasts in block 2, as the dependent variable and examines only the treatments in which subjects received feedback about both the strategies and outcomes of their peers. The first two columns of Table \ref{tab:corr} repeat specifications from earlier tables. Similar to the specification in column \ref{tab:baseline}.C, column \ref{tab:corr}.A includes an indicator for competitive rewards, along with the controls for subjects' forecasts and luck in the first block.  After controlling for individuals' early decisions and outcomes, the coefficient estimate for 1(Competitive) is not statistically significant. Column \ref{tab:corr}.B includes a measure of the relative distance between a subject's score and the score of the worst-ranked peer in the group and its interaction with the indicator for competitive rewards, similar to the specification in column \ref{tab:scores}.C. In column \ref{tab:corr}.B, the coefficient estimate on the relative distance is negative and statistically significant ($p=0.047)$; holding all else fixed in the noncompetitive treatment, subjects adopt riskier strategies as the distance between their score and the leader's score increases.\footnote{In contrast, in Table \ref{tab:scores}, the distance to the highest scorer had no statistically significant effect under noncompetitive rewards.} However, there is no similar effect under competitive rewards ($p=0.396$ for the sum of coefficient estimates on RelDist and its interaction with 1(Competitive)).

Subjects in treatments with peer information about both strategies and scores can learn about not just their relative positions, but also about the overall relationship between evidence gathering and outcomes in this environment (cf. Conjecture \ref{c_both}). We capture the empirical relationship between strategies and score by calculating, for each group, the Spearman rank-order correlation coefficient of the number of forecasts and subjects' scores in the first block. The correlation coefficients ranged from $-0.89$ to 0.82 across the 31 groups that were presented with the combined information. 

The regression reported in column \ref{tab:corr}.C includes the correlation coefficient based on the strategy and score data displayed to subjects after block 1 and its interaction with the indicator for competitive incentives. The coefficient estimate for the correlation variable is positive and statistically significant ($p=0.002$). We interpret the results by considering three cases: correlation coefficients of $-1$, 0, and 1. A correlation coefficient of $-1$ implies that scores increase with evidence gathering and conveys a noiseless and theoretically incorrect assessment of the relationship. Interpreting the estimates in column \ref{tab:corr}.C, a subject in the presence of this precise but incorrect information would take similar risk under competitive and noncompetitive rewards. In contrast, a correlation coefficient of 1 implies that scores decrease with evidence gathering and conveys a noiseless and theoretically correct assessment of the relationship. Presented with this information, our estimates suggest that subjects would take more risk when they face competitive compensation, relative to their risk taking under noncompetitive rewards ($p=0.022$). A correlation coefficient of 0 implies no observable, linear relationship between evidence gathering and scores. With noisy information, our subjects again would take more risk when they face competitive rewards as compared to noncompetitive compensation ($p=0.028$).

Keeping incentives fixed, the results in column \ref{tab:corr}.C imply that, under noncompetitive incentives, subjects exhibit a correct and statistically significant reaction to the correlation. The stronger the correlation between risk and rewards they observe in their group, the more risk they take. Under competitive incentives, however, the effect of the correlation on risk taking disappears. We summarize this finding as follows.

\begin{result}
\label{res6}
Under noncompetitive incentives, subjects gather less evidence when the observed correlation between risk and rewards is stronger. However, no similar effect is found for the setting with competitive incentives.
\end{result}

Column \ref{tab:corr}.D reports a demanding specification that includes the relative distance and correlation variables and their respective interactions with the indicator for competitive rewards. The coefficient estimates for the correlation variable and its interaction with the indicator for competitive incentives are similar in sign, magnitude, and statistical significance to the results in column \ref{tab:corr}.C. The coefficient estimates for the relative distance are not statistically significant for both reward schemes.

Column \ref{tab:corr}.E includes the measure of the correlation between evidence gathering and scores and the number of forecasts made by the best- and worst-scoring members of each group in block 1. In this specification, a higher highest score in the previous block is associated with more risk taking, although the effect is marginally significant ($p=0.052$). Coefficient estimates on the correlation measure and its interaction with the indicator for competitive rewards are both statistically significant ($p<0.001$ and $p=0.023$, respectively). We can again interpret the coefficients to compare competitive and noncompetitive incentives treatments: When the correlation coefficient is $-1$ (an incorrect characterization of the relationship between risk and score) or 0 (no obvious relationship), there is more risk taking in the competitive compensation treatments relative to the noncompetitive treatments; however, when the correlation coefficient is 1 (a correct characterization of the relationship), competitive incentives are associated with less risk taking.

Similar to the within-subject analysis discussed briefly at the end of Section \ref{sec:scoresonly}, we examine the difference between a subject's risk taking in the second and first block combining data from the treatments with no peer information and treatments with information on peers' strategies and scores. Regressing the change in risk taking on the indicators for peer information, competitive rewards and their interactions, along with the controls for first block risk taking and luck, we ask how a subject's risk taking is affected by learning not just that he or she has either the highest or lowest score but also \textit{why} they lead or lag their group. In contrast to the treatments with information about peers' scores only, the risk taking of subjects who learn that they are the highest- or lowest-scorers in their group does not change in a statistically significant way. Thus, there is no evidence supporting Conjecture \ref{c_both}. This is consistent with the results in columns \ref{tab:corr}.B and \ref{tab:corr}.D, where the risk taking in treatments with feedback about peers' strategies and scores appears unaffected by subjects' relative position in the group.

In the text responses to the follow-up questions, 70\% of the subjects claimed that their strategy changed over the rounds of the experiment (question (ii)). Approximately 35\% of the subjects wrote that these changes occurred after observing peers' strategies and scores, with 8\% (15\%) of the subjects stating that they switched to a safer (riskier) strategy (question (iii)). However, when asked directly, nearly no subject indicated that he or she imitated other subjects' strategies (question (iv)). This pattern of responses did not vary between the competitive and noncompetitive treatments. 

\subsection{Decisions in blocks 3 and 4}
\label{sec:blocks34}

Until now, our analysis has focused on behavior in block 2, using block 1 data to control for subjects' choices and outcomes in the first stage of the experiment. Recall that block 1 was identical across treatments, and the first treatment-specific feedback was given between blocks 1 and 2; therefore, the analysis of behavior in block 2 (conditional on block 1) allows us to estimate causal effects of feedback. In the final two blocks---block 3 and block 4---we are unable to identify any such causal relationship due to the endogeneity of feedback and score histories from earlier blocks. That said, even without the econometric identification of causal effects, exploring the longer-run correlations between evidence gathering, feedback and incentive schemes may be interesting from a practical point of view.

\begin{table}[tbp]
\begin{center}
\caption{Trends in evidence gathering, blocks 2-4}
{\small
{
\def\sym#1{\ifmmode^{#1}\else\(^{#1}\)\fi}
\begin{tabular}{l*{3}{c}}
\hline\hline
\multicolumn{4}{l}{Dependent variable: \textit{\# of forecasts}}\\ &\multicolumn{1}{c}{\ref{tab:trends}.A}&\multicolumn{1}{c}{\ref{tab:trends}.B}&\multicolumn{1}{c}{\ref{tab:trends}.C}\\
\hline
Block       &       -0.60\sym{***}&       -0.76\sym{***}&       -0.81\sym{***}\\
            &      (0.11)         &      (0.16)         &      (0.22)         \\
Block$\times$1(Competitive)&                     &        0.32\sym{*}  &        0.09             \\
            &                     &      (0.17)         &     (0.27)          \\
Block$\times$1(Peer information: strategies only)    &                     &                     &        0.01         \\
            &                     &                     &      (0.30)         \\
Block$\times$1(Peer information: scores only)    &                     &                     &       -0.03         \\
            &                     &                     &      (0.39)         \\
Block$\times$1(Peer information: strategies \& scores)    &                     &                     &        0.13         \\
            &                     &                     &      (0.30)         \\
Block$\times$1(Competitive)   &                     &                     &        0.25         \\
\hspace*{2em}$\times$1(Peer information: strategies only)            &                     &                     &      (0.29)         \\
Block$\times$1(Competitive)   &                     &                     &        0.46         \\
\hspace*{2em}$\times$1(Peer information: scores only)            &                     &                     &      (0.39)         \\
Block$\times$1(Competitive)   &                     &                     &        0.53\sym{**} \\
\hspace*{2em}$\times$1(Peer information: strategies \& scores)            &                     &                     &      (0.26)         \\
Constant      &       13.92\sym{***}&       13.93\sym{***}&       13.93\sym{***}\\
            &      (0.36)         &      (0.36)         &      (0.36)         \\
\hline
\(N\)       &        1200         &        1200         &        1200         \\
Pseudo $R^2$ &      0.002          &    0.004             &  0.006               \\
\hline\hline
\end{tabular}
}
\captionsetup{font=small}
\caption*{Note: Pooled Tobit regression results using data from all treatments in blocks 2-4. Standard errors, clustered by group, are reported in parentheses; there are 152 clusters in each regression. Group size is five in all treatments except those treatments without feedback, where the group size is one because subjects were not exposed to information about their peers. *, ** and *** indicate statistical significance at the 10\%, 5\% and 1\% levels, respectively.}
\label{tab:trends}
}
\end{center}
\end{table}

\begin{table}[tbp]
	\begin{center}
		\caption{Dynamics of evidence gathering, blocks 3 and 4}
		{\small
			{
				\def\sym#1{\ifmmode^{#1}\else\(^{#1}\)\fi}
				\begin{tabular}{l*{5}{c}}
					\hline\hline
					\multicolumn{6}{l}{Dependent variable: \textit{\# of forecasts}}\\ &\multicolumn{1}{c}{\ref{tab:dynamics}.A}&\multicolumn{1}{c}{\ref{tab:dynamics}.B}&\multicolumn{1}{c}{\ref{tab:dynamics}.C}&\multicolumn{1}{c}{\ref{tab:dynamics}.D}&\multicolumn{1}{c}{\ref{tab:dynamics}.E}\\
					\hline
					1(Competitive)  &        1.07\sym{*}  &        0.87  &        0.82         &       0.22         &       0.22         \\
					&      (0.59)         &      (0.52)         &      (0.51)         &      (0.80)         &      (0.80)         \\
					1(Peer information)    &                     &                     &                     &        0.67         &                     \\
					&                     &                     &                     &      (0.72)         &                     \\
					1(Competitive)$\times$1(Peer information) &                     &                     &                     &        0.75         &                     \\
					&                     &                     &                     &      (1.00)         &                     \\
					1(Peer information: strategies only)&                     &                     &                     &                     &       0.13         \\
					&                     &                     &                     &                     &      (0.79)         \\
					1(Peer information: scores only)&                     &                     &                     &                     &        0.79        \\
					&                     &                     &                     &                     &      (1.28)         \\
					1(Peer information: strategies &                     &                     &                     &                     &        0.86         \\
					\hspace*{2em}\& scores)            &                     &                     &                     &                     &      (0.84)         \\
					1(Competitive)$\times$1(Peer &                     &                     &                     &                     &        0.62         \\
					\hspace*{2em}information: strategies only)            &                     &                     &                     &                     &      (1.20)         \\
					1(Competitive)$\times$1(Peer &                     &                     &                     &                     &        0.66         \\
					\hspace*{2em}information: strategies only)            &                     &                     &                     &                     &      (1.70)         \\
					1(Competitive)$\times$1(Peer &                     &                     &                     &                     &       0.90         \\
					\hspace*{2em}information: strategies \& scores)            &                     &                     &                     &                     &      (1.13)         \\
					\# of forecasts in block 1&                     &        0.66\sym{***}&        0.66\sym{***}&        0.67\sym{***}&        0.67\sym{***}\\
					&                     &      (0.07)         &      (0.07)         &      (0.07)         &      (0.07)         \\
					Luck in block $t-1$    &                     &                     &        0.21\sym{***}&        0.21\sym{***}&        0.21\sym{***}\\
					&                     &                     &      (0.05)         &      (0.05)         &      (0.05)         \\
					Constant      &       11.27\sym{***}&        2.71\sym{***}         &       2.86\sym{***}         &        2.28\sym{**}         &        2.24\sym{**}         \\
					&      (0.45)         &      (0.91)         &      (0.91)         &      (1.00)         &      (1.00)         \\
					\hline
					\(N\)       &        800         &        800         &        800         &        800         &        800         \\
					Pseudo $R^2$       &   0.002   &  0.043     &   0.048     &  0.050    &   0.051   \\
					\hline\hline
				\end{tabular}
			}
			\captionsetup{font=small}
			\caption*{Note: Pooled Tobit regressions using data from all treatments in blocks 3 and 4. Standard errors, clustered by group, are reported in parentheses; there are 152 clusters in each regression. Group size is five in all treatments except those treatments without feedback, where the group size is one because subjects were not exposed to information about their peers. *, ** and *** indicate statistical significance at the 10\%, 5\% and 1\% levels, respectively.}
			\label{tab:dynamics}
		}
	\end{center}
\end{table}

Panels C and D of Table \ref{tab:sumstat} present summary statistics by treatment for the third and fourth blocks. On average, subjects continue to engage in less evidence gathering under competitive rewards in those blocks. The summary statistics in Table \ref{tab:sumstat} also suggest that evidence gathering increases over time. To explore the time trends, we use data from blocks 2--4 as a panel to estimate pooled Tobit specifications and present the results in Table \ref{tab:trends}. The coefficient estimates in column \ref{tab:trends}.A suggest a significant positive time trend; average evidence gathering is increasing over the experiment. Results in column \ref{tab:trends}.B suggest that this upward trend is especially strong under noncompetitive incentives ($p=0.001$). Finally, the specification in column \ref{tab:trends}.C allows us to measure the time trends separately by treatment. Wald tests suggest that the trend is statistically significant at $p<0.05$ in each treatment, except under competitive incentives with information on scores ($p=0.307$) and full feedback ($p=0.093$). Together, the results suggest that costly evidence gathering increases over the experiment, but increases more slowly when subjects face competitive incentives and have information on their relative performance.

%In (3), check the sums of coefficients on block and block$\_$t for each treatment t, to see if the trend disappears in some treatment (most likely, 13 and 14): t2 $p=0.001$, t3 $p=0.021$, t4 $p=0.003$, t11 $p=0.0004$, t12 $p=0.023$, t13 $p=0.307$, t14 $p=0.093$.

We also examine subjects' behavior in blocks 3 and 4 after accounting for their decisions in block 1 and luck in the previous block, similar to the analysis in Section \ref{sec:main_results}. Table \ref{tab:dynamics} presents the results of pooled Tobit specifications parallel to those in Table \ref{tab:baseline} using panel data from blocks 3 and 4.\footnote{We exclude block 2 from these panels because decisions in block 2 have been analyzed in detail above and because decisions in block 2 are influenced by block 1, a substantially different setting than the subsequent blocks.} As a dynamic control, we include the lagged measure of luck. The overall results are very similar to those in Table \ref{tab:baseline}: there is less evidence gathering under competitive incentives without additional controls (column \ref{tab:dynamics}.A), and the coefficient estimate is similar in magnitude. However, more subtle patterns no longer hold. The coefficient estimate on the indicator for competitive incentives is only marginally significant in column \ref{tab:dynamics}.B when we control for decisions in block 1 ($p=0.092$), and it becomes insignificant in \ref{tab:dynamics}.C when we additionally control for prior luck ($p=0.110$). Furthermore, the coefficient estimate on the indicator for competitive incentives in the absence of peer information is not statistically significant ($p=0.783$), and neither is the effect of feedback under competitive incentives ($p=0.108$, column \ref{tab:dynamics}.D).  Consistent with part (a) of Result \ref{res1}, there is no robust relationship between competitive incentives and the average level of evidence gathering in later blocks; moreover, parts (b) and (c) of the result---an increase in evidence gathering under competition without and with feedback---no longer hold. However, there still is a positive and significant effect of feedback on the number of forecasts---a negative effect on evidence gathering---in the presence of competitive incentives ($p=0.043$, column \ref{tab:dynamics}.D).

The results presented in column \ref{tab:dynamics}.E allow us to also consider different types of feedback. We find that the relationship between incentives, feedback and risk taking is driven primarily by the treatment with information about both peers' strategies and scores ($p=0.023$ for the sum of coefficients on combined feedback and its interaction with competitive incentives). This finding supports the robustness of Result \ref{res2}(c) on the effect of combined feedback; other parts of Result \ref{res2} do not persist in later blocks.

We can also use data from blocks 3 and 4 to reproduce the analysis for the treatments with information on scores only (as in Section \ref{sec:scoresonly}) and with information on strategies and scores (as in Section \ref{sec:strategies_scores}). The corresponding regression results are presented in Tables \ref{tab:scores34} and \ref{tab:corr34} in  \ref{app_34}. Table \ref{tab:scores34} is parallel to Table \ref{tab:scores}, and Table \ref{tab:corr34} is parallel to Table \ref{tab:corr}. Instead of controls for luck, distances to the highest and lowest score, and other variables based on block 1 data, we now create dynamic controls using data from the previous block. By construction, these analyses suffer from endogenity and, as a result, they do not lead to any causal estimates or claims. For the treatments with information on scores only, we find no correlation between the distance measures and evidence gathering in later blocks. In the treatments with full feedback, we find statistically significant associations similar to those identified causally in Table \ref{tab:corr}. Specifically, the empirical correlation between strategies and scores in the group in the previous block is associated negatively with evidence gathering under noncompetitive rewards, but there is no effect under competitive rewards. Also, the number of forecasts by the subjects with the highest score and lowest score are associated negatively with evidence gathering---only the former effect is statistically significant in block 2. 

Overall, our analysis of outcomes from blocks 3 and 4 suggests that some of the causal relationships identified in block 2 persist as associations in later blocks. First, combined feedback about peers' strategies and outcomes is associated with less evidence gathering under competitive incentives; second, there is a robust effect of empirical correlation between risk taking and scores, whereby subjects update their strategies in the direction of the observed relationship under noncompetitive incentives, but not under competitive incentives. Of course, without clean econometric identification, any associations beyond the second block should be interpreted cautiously.

\section{Discussion and conclusion}
\label{sec:conclusion}

In this paper, we explore the effect of incentives and peer information on evidence gathering in an experimental forecasting task. Our results suggest that the relationship between incentives and risk taking is nuanced and depends critically on both the availability and content of peer-related information. Specifically, in the absence of peer information, subjects may engage in more risk-reducing evidence gathering in a setting with competitive incentives than with noncompetitive incentives; our estimates are large in magnitude, albeit marginally significant ($p<0.10$). At the same time, we find strong statistical evidence that subjects gather less evidence, i.e., take more risk, under competitive incentives in the presence of peer feedback. 

In our experimental setting, the more cautious behavior of subjects under competition without peer feedback may be attributed to the type of the environment, underconfidence, and the fear of being last in the three-level prize structure: Without knowing what their peers are doing, subjects may be underconfident when completing difficult tasks due to ``reference group neglect'' \citep{Moore-Cain:2007} and, combined with the fear of being last \citep{Dutcher-et-al:2015}, subjects may be more inclined to make low-risk choices. %This baseline result makes the effect of competitive incentives on evidence gathering in the presence of peer information even more striking as it reverses the baseline predisposition of subjects to respond to competition cautiously in our setting.

In the presence of peer feedback, regardless of its type, most subjects observe that their strategies and/or outcomes are not that far off from their peers'. Peer information, therefore, may reduce subjects' underconfidence and their concerns about a last-place finish. Overall, feedback about others may serve as a simple mechanism that restores ``normal'' attitudes to competition and, as a result, leads to a decline in evidence gathering---and more risk taking---under competitive incentives.

The effect of incentives on evidence gathering does not seem to depend systematically on the type of peer information (as long as it is present); however, holding fixed the incentives, different types of feedback are associated with different amounts of evidence gathering. 

Under noncompetitive incentives and compared to the no-feedback baseline, subjects take less risk when provided with information about peers' strategies or outcomes separately, but not when provided with both types of information. In the cases of information on strategies or outcomes only, subjects do not seem to be reacting to the content of the feedback in any imitative manner. With feedback on strategies, subjects do not imitate the highest, lowest or median levels of risk taking, nor do they converge to the mean. With feedback on scores, subjects' behavior is not affected by the distance to their group's highest or lowest score. Both types of partial information are not particularly useful and may reaffirm subjects' impression that the game is complex and noisy; in the case of noncompetitive incentives, partial information reduces risk taking. When the combined feedback on peers' strategies and scores is provided, subjects react correctly to the correlation between the observed risk taking and scores and, perhaps as a result, do not reduce their risk taking relative to the no-feedback case. 

Under competitive incentives, there is no effect of peer information on risk taking when only peers' strategies are disclosed. With feedback on peers' outcomes, there is some evidence that subjects react in an expected way to the payoff-relevant distances between their score and the scores of the top and bottom performers, although these effects are noisy. At the same time, subjects facing competitive rewards do not react to information revealing the correlation between strategies and outcomes.

In all treatments, subjects react positively to luck, defined in this study as the difference between a subject's score and the theoretically expected score given his or her strategy in the first block. The effect of luck is asymmetric: If a subject takes a lot of risk and gets lucky, there is little room to increase risk taking further; if that same subject is unlucky, he or she can take substantially less risk. Similarly, if a subject takes very little risk and is unlucky, there is no room to reduce risk taking; if the same subject is lucky, there is room to take much more risk. Of course, the difference is that the variance of score is much higher for high levels of risk. That is, subjects are much more likely to be lucky or unlucky (and the amount of good or bad luck is likely to be larger) if they take more risk. Therefore, the effect of prior luck on risk taking is driven primarily by the reduction in risk taking by the unlucky high-risk subjects.

Some insights into the effects of peer information can be gained by analyzing subjects' responses to the open-ended questions at the end of the experiment.  A substantial number of subjects state that they changed behavior after observing their peers. As one would expect, this response is most common in treatments with combined information and least common when subjects observe only their peers' strategies. At the same time, almost no subjects claim that they imitate specific peers. Taken along with the empirical results, subjects' written responses suggest that peer information is interpreted in an aggregate, inferential manner, as opposed to being a path to simple imitation.

%We conclude that although incentives can be a powerful driver of risk-taking behavior, peer information is also important---it is the interaction of the two that leads to the highest levels of risk taking in our setting. More broadly, our results suggest that excessive risk taking in, for example, the financial sector may be the combined result of incentives, culture, norms, and the feedback received by the decision-makers. The nature of the information clearly matters, as it shapes decision-makers' understanding of the sources of success in a noisy environment. Of course, selection may distort the availability of information in the field and spur increased risk taking. In many settings, success stories are more likely to be propagated; in high-variance environments in which success has a large random component, this selection may lead to excessive risk taking.

The common view in the literature---as well as the prevailing opinion of nonacademic commentators---has been that competitive incentives lead to more risk taking. Therefore, more generally, our findings suggest that the common wisdom is perhaps less universal than previously thought. Moreover, we can offer no general prescription for an ideal feedback or rewards scheme, as there is no universally optimal amount of risk for all settings. Our results do imply, however, that the principal can manipulate the level of risk to some extent through two levers: the rewards structure and the feedback policy. Consider, for example, a setting where agents are engaged in reliability or quality control tasks, such as software testing or semiconductor chip testing. A principal who wants to increase production volume (at the expense of more variability in quality) can facilitate the flow of information about strategies and outcomes among workers. In contrast, a principal seeking to raise average quality and reduce variability (at the expense of volume) would want to limit the amount of information that workers can exchange. 

Our results also show that alternative risk-taking tasks, including our task, can help uncover new and unexpected effects. Indeed, a trade-off between risk and volume can be introduced in a variety of ways. While the investment task used in most of the literature is simple and analytically tractable, it is quite different from the scenarios where a fixed resource, such as time, is allocated among multiple information gathering decisions (e.g., software testing, quality control, journalistic or criminal investigation, or exam taking). In the context of R\&D, risk taking can be implemented dynamically as a stopping problem \citep[e.g.,][]{Seel-Strack:2013} or statically in the form of a flexible mean-preserving spread added to the agents' initial positions \citep[e.g.,][]{Ke-et-al:2021}. These new tasks have counterparts in the real world and are worth pursuing experimentally, if anything, as a robustness check for the existing results based on the investment task.

%No universal effect of competitive incentives on risk taking: less risk taking without feedback, more risk taking with feedback (as compared to noncompetitive). Feedback matters even when it contains no helpful information. Salience of the group + relative incentives -> more risk taking? 

%What about responses to questions (i) and (ii) in the treatment without feedback? If the above conjecture is correct, should see more people saying they don't know what to do, have no strategy, switch to safer strategy. Nope, that's not the case. The proportion of people saying they have no strategy is the same across treatments.

%Some ideas: Completeness of feedback is important. Imagine selective feedback where only successful people tell their stories. Can we run counterfactual regression to see what happens to risk taking if all peers' scores are high? Extension: allow people to decide if they want to provide feedback (or to receive feedback) Problem with our design: risk taking is not ``bad,'' most subjects take too little risk. Would be nice to have a setting where optimal amount of risk is low (maybe, for society, but not individually; negative externality).

\pagebreak

\section*{Declarations}
\begin{itemize}

\item The authors have no relevant financial or non-financial interests to disclose.

\item This project was reviewed and approved by the Florida State University Human Subjects Committee, approval number 2013.9805. 

\end{itemize}

\pagebreak
\bibliographystyle{aea}

\pagebreak

\appendix

\renewcommand{\thesection}{Appendix \Alph{section}}
\renewcommand{\thesubsection}{A\arabic{subsection}}

\singlespacing

%\section*{Appendix}

\section{Probability matching}
\label{app_PM}

\cite{Vulkan:2000} provides a recent survey of the literature on probability matching from an economist's perspective. A typical experiment on probability matching involves subjects observing a random sequence of two signals (e.g., red and green lights) that they are told are independent draws in which the red signal appears with some probability $p$. Having observed a sufficiently long sequence, subjects are then asked to predict what the next signal will be. Clearly, red is the optimal prediction if and only if red light was observed more than half of the time. Instead, many subjects randomize their predictions and predict red with a frequency close to $p$. This suboptimal randomizing behavior is robust to learning, although it can be partially mitigated through guided deliberation and advice \citep{Koehler-James:2009}. Recently, \cite{Gaissmaier-Schooler:2008} proposed that probability matching serves as an evolutionarily ``smart'' heuristic whereby individuals look for patterns and serial dependence in naturally occurring stimuli.

Our forecasting task differs in two significant ways from the simple settings in which probability matching has been studied previously. First, it includes variation in the intensity of the signal; our subjects see different numbers of red and green cards revealed. Second, there is variation in the amount of noise; our subjects observe the differences in the numbers of red and green cards revealed and the number of hidden cards. Suppose that a subject flips $n$ out of $M$ total cards, $r$ of which turn out to be red and $g$ are green. The larger the difference $|r-g|$ and the smaller the number of hidden cards $M-n$, the more likely it is that the majority color forecast will be correct. Thus, while a payoff maximizer will choose the majority color all of the time, a probability matcher will choose the majority color with a probability that is increasing in $|r-g|$ and $n$. Testing this formally in a probit regression, we find negative and statistically significant effects of $|r-g|$ and $n$ on the probability of a nonmajority guess. In our experiment, probability matching is present across all of the compensation and peer information conditions.

\begin{figure}[htbp]
\includegraphics[width=3.0in]{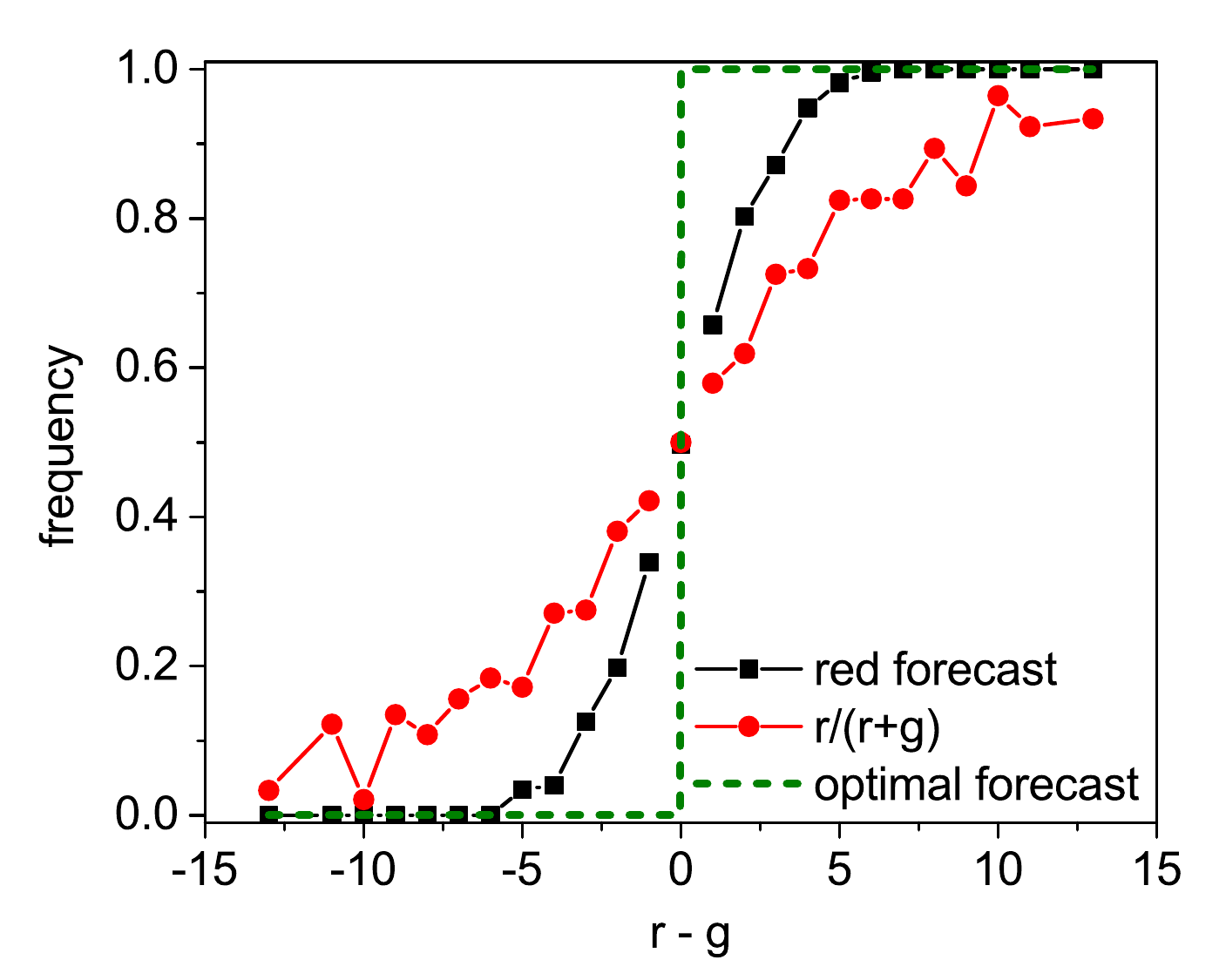}
\includegraphics[width=3.0in]{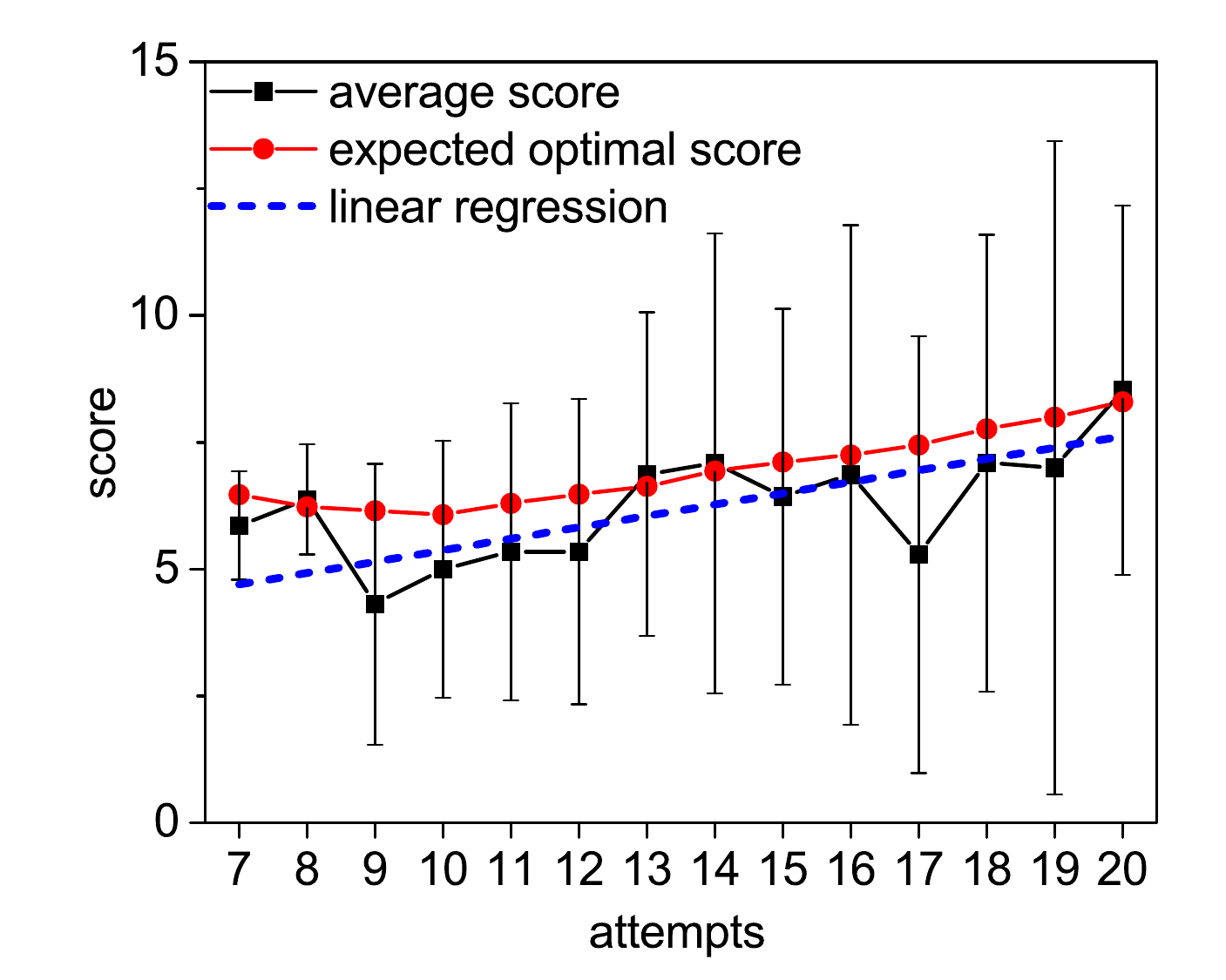}
\caption{\textit{Left}: The average frequency of forecasting red color as a function of $r-g$ (squares); average ratio $r/(r+g)$ as a function of $r-g$ (circles); the optimal forecasting strategy (dashed line). \textit{Right}: Average score in block 1 as a function of the number of attempts (squares), with error bars showing one standard deviation above and below the average score; average expected score in block 1 if all forecasts were optimal (circles); regression line of average score on attempts, including an intercept (dashed).}
\label{fig:prob_matching}
\end{figure}

In the left panel in Figure \ref{fig:prob_matching}, the squares plot the observed average frequency of red forecasts against the difference between the number of red and green cards, $r-g$. The optimal forecasting strategy is shown by a dashed line. The circles plot the average ratio of red cards to the total number of cards, $\frac{r}{r+g}$, against each value of $r-g$ observed in the experiment. The ratio $\frac{r}{r+g}$ represents the behavior of a ``pure'' probability matcher. The observed average behavior lies between pure probability matching and the optimal behavior, suggesting that probability matching varies across subjects and with the level of noise, as observations for each $r-g$ are averaged over different values of $n$. As expected, the frequency of forecasting red increases monotonically in $r-g$, starting at zero for $r-g\le -6$ and reaching one at $r-g\ge 7$. 

The pattern of choices observed in the experiment is consistent with random choice models, such as the Quantal Response Equilibrium (QRE) model introduced by \cite{McKelvey-Palfrey:1995}. In the QRE framework, it is assumed that instead of choosing a utility-maximizing strategy, a subject chooses a strategy $s$ with probability $p(s)=\frac{\exp[\lambda u(s)]}{\sum_{s'\in S}\exp[\lambda u(s')]}$, where $S$ is the set of possible strategies and $u(s)$ is the expected utility of strategy $s$. Parameter $\lambda$ represents the (inverse) level of noise (or intensity of errors in decision-making), where $\lambda\to 0$ corresponds to completely random behavior and $\lambda\to\infty$ corresponds to fully rational behavior. In our setting, there are two strategies: choose red (R) or choose green (G). Let $p_{n,r}$ denote the probability that red is the majority color given that $n$ cards have been flipped and $r$ of them are red (see equation (\ref{pnr})). The expected utility of choosing R is $u(R)=2p_{n,r}-1$, and the expected utility of choosing G is $u(G)=2(1-p_{n,r})-1$. Therefore, the logistic probability of choosing red is $p(R) = \frac{1}{1+\exp[2\lambda(1-2p_{n,r})]}$. Maximum likelihood estimation using choice data from the experiment yields $\lambda=1.4$ (standard error of 0.02). There is little variation in $\lambda$ across blocks and treatments, suggesting that the amount of noise in the random choice model does not depend on the compensation or feedback conditions.

Given the presence of suboptimal probability matching behavior, we consider the extent to which such decisions distort the evidence gathering incentives in the experiment. In particular, we explore how probability matching affects the trade-off between risk and expected returns predicted by the theory in Section \ref{sec:predictions}, which assumes that subjects always forecast the majority color. The squares in the right panel of Figure \ref{fig:prob_matching} plot subjects' average score in the first block against the number of forecasting attempts. The error bars around the squares show one standard deviation above and below the mean. The number of forecasts ranges from 7 (the safest strategy, flipping 15 cards in each period except the last one) to 20 (the riskiest strategy, flipping 5 cards in each period). The dashed line is a linear regression line for the average score as a function of the number of forecasts, including an intercept. The slope is positive and statistically significant at conventional levels. The circles show the expected score that a subject would have received if he or she always followed the majority guessing strategy. Due to suboptimal decision-making, the average score in the experiment is approximately 0.7 lower than the expected optimal score, and the difference is statistically significant. However, the observed dependence of score on the number of forecasting attempts is essentially a parallel shift down from what fully rational theory predicts, and the trade-off between risk and returns is preserved despite the presence of probability matching.

\newpage
\section{Additional tables for blocks 3 and 4}
\label{app_34}

\begin{table}[h]
	\begin{center}
		\caption{Distance to the best- and worst-scoring peers, blocks 3 and 4}
		{\small
			\def\sym#1{\ifmmode^{#1}\else\(^{#1}\)\fi}
			\begin{tabular}{lccc}
				\hline\hline
				\multicolumn{3}{l}{Dependent variable: \textit{\#\ of forecasts}} \\
				&       \ref{tab:scores34}.A    &         \ref{tab:scores34}.B    &         \ref{tab:scores34}.C   \\
				\hline
				1(Competitive)  &       0.81    &      -0.09    &   0.22   \\
				&      (1.16)    &     (1.63)    &     (1.81)   \\
				Distance to the group's highest 		&        0.49\sym{*}    &                &         \\
				\hspace*{2em}  score in block $t-1$                 &      (0.26)    &                &        \\
				1(Competitive)$\times$Distance to the &        -0.08 &                &       \\
				\hspace*{2em} group's highest score in block $t-1$         &      (0.25)    &                &        \\
				Distance to the group's lowest		&                 &       -0.13    &          \\
				\hspace*{2em}  score in block $t-1$                    &                 &     (0.44)    &        \\
				1(Competitive)$\times$Distance to the &                 &      0.29    &         \\
				\hspace*{2em} group's lowest score in block 1         &                 &     (0.27)    &        \\
				RelDist in block $t-1$ &                 &          &  -2.51       \\
				&                 &         &  (2.23)      \\
				1(Competitive)$\times$RelDist in block $t-1$ &                 &          &  1.00       \\
				&                 &         &  (1.42)      \\
				\#\ of forecasts in block 1      &        0.81\sym{***} &       0.77\sym{***} &       0.79\sym{***}\\
				&      (0.12)    &     (0.13)    &     (0.12)   \\
				Luck in block $t-1$				&        0.60\sym{**} &       0.21  &       0.46 \\
				&     (0.26)    &     (0.42)    &     (0.31)   \\
				Constant            &        -0.00    &       2.10    &       2.98   \\
				&      (1.58)    &     (2.40)    &     (2.98)   \\
				\hline
				$N$                   &           80    &          80    &          80   \\
				Pseudo $R^2$                &        0.066    &       0.057    &       0.058   \\
				\hline\hline
			\end{tabular}
			\captionsetup{font=small}
			\caption*{Note: Pooled Tobit regression results using data from treatments in which subjects received information about peers' scores only, blocks 3 and 4. Group-level clustered standard errors are reported in parentheses; there are 16 clusters in each regression. *,  ** and *** indicate statistical significance at the 10\%, 5\% and 1\% levels, respectively.}
			\label{tab:scores34}
		}
	\end{center}
\end{table}

\begin{table}[tbp]
	\begin{center}
		\caption{Strategies and scores, blocks 3 and 4}
		{\small
			\def\sym#1{\ifmmode^{#1}\else\(^{#1}\)\fi}
			\begin{tabular}{lccccc}
				\hline\hline
				\multicolumn{3}{l}{Dependent variable: \textit{\#\ of forecasts}} &       &     \\
				&        \ref{tab:corr34}.A     &        \ref{tab:corr34}.B     &        \ref{tab:corr34}.C    &         \ref{tab:corr34}.D   &          \ref{tab:corr34}.E   \\
				\hline
				1(Competitive)            					&      1.12     &      0.46     &      0.69  &      0.01   &        0.38 \\
				&    (0.82)     &    (0.83)     &    (0.79)    &     (0.85)   &      (0.58)   \\
				RelDist in block $t-1$ 		  				&                &      -0.36    &               &       -0.14   &                \\
				&                &    (1.08)     &               &     (1.20)   &                \\
				1(Competitive)$\times$RelDist in block $t-1$ 		&                &      1.32     &               &       1.36   &                \\
				&                &    (1.03)     &               &     (1.05)   &                \\
				Correlation of forecasts and	         			&                &                &     2.46\sym{***} &      2.47\sym{***} &       1.02\\
				\hspace*{2em}   scores in block $t-1$                  	&                &                &    (0.80)    &     (0.81)   &      (0.82)   \\
				1(Competitive)$\times$Correlation of forecasts     	&                &                &      -1.07    &      -1.03   &        0.50 \\
				\hspace*{2em}  and scores in block $t-1$                    &                &                &    (1.10)    &     (1.16)   &      (0.79)   \\
				\#\ of forecasts by subject with highest	  		&                &                &               &               &        0.38\sym{***}  \\
				\hspace*{2em}  score in block $t-1$                       	&                &                &               &               &      (0.09)   \\
				\#\ of forecasts by subject with lowest	   		&                &                &               &               &        0.27\sym{***}   \\
				\hspace*{2em}  score in block $t-1$                        	&                &                &               &               &      (0.09)   \\
				\#\ of forecasts in block 1        				&      0.61\sym{***}  &      0.60\sym{***}  &      0.59\sym{***} &       0.58\sym{***}&        0.47\sym{***}\\
				&    (0.11)     &    (0.11)     &    (0.10)    &     (0.10)   &      (0.09)   \\
				Luck in block $t-1$  						&      0.06  &      0.04  &      0.05  &       -0.00  &        0.05 \\
				&    (0.06)     &    (0.14)     &    (0.06)    &     (0.15)   &      (0.05)   \\
				Constant            						&      3.85\sym{**}   &      4.11\sym{***}    &      4.28\sym{***}  &       4.46\sym{***} &       -2.99\sym{**}   \\
				&    (1.54)     &    (1.51)     &    (1.40)    &     (1.48)   &      (1.31)   \\
				\hline
				$N$                     &        310     &        310     &        310    &         310   &          310   \\
				Pseudo $R^2$          &    0.041       &   0.042        &   0.050       &   0.051       &  0.079       \\ 
				\hline\hline
			\end{tabular}
			\captionsetup{font=small}
			\caption*{Note: Pooled Tobit regression results using data from treatments in which subjects received information about peers' strategies and scores, blocks 3 and 4. Group-level clustered standard errors are reported in parentheses; there are 31 clusters in each regression. *, ** and *** indicate statistical significance at the 10\%, 5\% and 1\% levels, respectively.}
			\label{tab:corr34}
		}
	\end{center}
\end{table}

\newpage
\section{Experimental instructions}
\label{sec:instructions}

\parindent 0pt

\begin{center}\textbf{Instructions for Part 1a}\end{center}

In this task, your decision will generate a set of 20 choices between a lottery that will be referred to as ``Urn A'' and sure amounts of money. After you have made your decision, one of the 20 choices will be selected randomly and played. 

\bigskip

\textbf{Urn A contains 20 balls, 10 of which are green and 10 are red}.

\bigskip

If your preference in the choice that turns out to be actually played is Urn A, you earnings will depend on your guess about the color of a ball randomly drawn from Urn A. If you guess the color correctly, you will earn \$2.00. If you guess the color incorrectly, you will earn zero.

\bigskip

If your preference in the choice that turns out to be actually played is a sure amount of money, you will earn that amount of money.

\bigskip

On the screen, you can see all the 20 choices. This is practice screen, and all buttons are now inactive. Urn A is on the left, and the sure amounts of money ranging from \$0.10 to \$2.00 are on the right. Notice that the sure amounts increase from top to bottom. Thus, you should only decide on a line where you would like to SWITCH from preferring Urn A to preferring a sure amount. 

\bigskip

When you click on the corresponding SWITCH HERE button, Urn A will be your choice everywhere above that line, and a sure amount of money will be your choice everywhere below that line. All the 20 choices that you generate will be highlighted. If you want to change your decision, simply click on another SWITCH HERE button. When you are ready to finalize your decision, click SUBMIT.

\bigskip

You will be informed about your earnings from this part of the experiment at the very end of the session today, after you have completed all parts of the experiment.

\bigskip

Are there any questions before you begin making your decisions?

\bigskip
 
\begin{center}\textbf{Instructions for Part 1b}\end{center}

In this task, your decision will generate a set of 20 choices between a lottery that will be referred to as ``Urn B'' and sure amounts of money. After you have made your decision, one of the 20 choices will be selected randomly and played. 

\bigskip

{\bf Urn B contains 20 balls that are either green or red. The exact numbers of green and red balls are unknown to you}.

\bigskip

If your preference in the choice that turns out to be actually played is Urn B, earnings will depend on your guess about the color of a ball randomly drawn from Urn B. If you guess the color correctly, you will earn \$2.00. If you guess the color incorrectly, you will earn zero.

\bigskip

If your preference in the choice that turns out to be actually played is a sure amount of money, you will earn that amount of money.

\bigskip

On the screen, you can see all the 20 choices. This is practice screen, and all buttons are now inactive. Urn B is on the left, and the sure amounts of money ranging from \$0.10 to \$2.00 are on the right. Notice that the sure amounts increase from top to bottom. Thus, you should only decide on a line where you would like to SWITCH from preferring Urn B to preferring a sure amount. 

\bigskip

When you click on the corresponding SWITCH HERE button, Urn B will be your choice everywhere above that line, and a sure amount of money will be your choice everywhere below that line. All the 20 choices that you generate will be highlighted. If you want to change your decision, simply click on another SWITCH HERE button. When you are ready to finalize your decision, click SUBMIT.

\bigskip

You will be informed about your earnings from this part of the experiment at the very end of the session today, after you have completed all parts of the experiment.

\bigskip

Are there any questions before you begin making your decisions?

\bigskip

\begin{center}\textbf{Instructions for Part 2}\end{center}

\underline{The scenario}

\bigskip

Imagine that you are a financial analyst hired by an investment company to make projections about the future performance of particular stocks. Your job is to assess whether a company's stock price next year will be HIGHER or LOWER than the current price. You are assigned one project at a time and are paid based on the volume and accuracy of your forecasts -- your employer rewards you for correct forecasts and punishes you for incorrect forecasts. 

\bigskip

You make your forecasts based on information that you can gather about the companies in question. Each piece of information provides a signal about the true direction of the stock price, suggesting that it is either going to be higher or lower next year. No single signal tells the full story; however, the more signals you observe the more confident you may be about whether the stock price will be higher or lower. 

\bigskip

Recall that you are paid based on the \textit{volume} and the \textit{accuracy} of your forecasts. The challenge that you face is that gathering lots of information can improve the accuracy of your forecasts, but means that you cannot do many assessments. Making forecasts with a small amount of information means that you can complete many projects, but these assessments may not be very accurate. 

\bigskip

\underline{Decision sequences}

\bigskip

This part of the experiment will consist of several decision sequences. At the end of the experiment, one of the sequences will be randomly chosen and your actual earnings will be based on that sequence.

\bigskip
\newpage

\underline{Sequence 1}

\bigskip

\textit{Task}

On the screen, you will be presented with 15 blank cards. Each card, when flipped over, is either GREEN or RED. The color of the card was determined randomly, and each of the two colors is equally likely.

\bigskip

Your task is to predict the \textit{majority color} of the 15 cards (hidden and revealed). At least 8 of the 15 cards are going to be of one color, GREEN or RED. If 8 or more cards are GREEN, then the majority color is GREEN; if 8 or more cards are RED, then the majority color is RED.

\bigskip

At the bottom of the screen, you can submit your forecast of whether the majority of the 15 cards (hidden and revealed) are GREEN or RED.

\bigskip

Before you make a forecast about the majority color, you should choose how many cards, \textbf{between 5 and 15}, you want to flip to reveal their color. After you make a forecast, all cards will be revealed and you will be informed whether your forecast was correct or not. You will then be given the next task.

\bigskip

In this sequence, you will be allowed to flip a \textbf{total of 100 cards}. How many cards you flip before each forecast is up to you. The counter in the upper part of the screen will tell you how many cards you have left to flip in this sequence. It starts with 100 cards and counts down. New tasks will be generated randomly until the counter of 100 cards runs out.

\bigskip

Note that you should reveal 5 or more cards for each forecast. Towards the end of the sequence when you nearly exhaust all of your 100 cards, you will not be allowed to reveal a number of cards such that the remaining number of cards is less than 5.

\bigskip

At the end of the sequence, you will be provided with a summary of your decisions and forecasts.

\bigskip

\textit{Score and payoff}

Your performance score in this sequence will be based on the net number of correct majority color forecasts calculated as

\bigskip

\textbf{Score = (\# of correct forecasts) - (\# of incorrect forecasts)}.

\bigskip

For example, if you make 3 correct forecasts and 1 incorrect forecast, your score would be $3-1=2$.

\bigskip

Your payoff from this sequence is \textbf{your score times \$1.50}. Recall that there will be several sequences, and only one of them will be chosen at the end of the experiment for your actual earnings. 

\bigskip

Are there any questions before you begin?

\bigskip

You will now start the actual decision rounds. Please do not communicate with other participants or look at their monitors. If you have a question or problem, from this point on please raise your hand and one of us will assist you in private.

\bigskip

\textit{The following instructions are shown on the screen in NONCOMPETITIVE treatments after the first block is completed.}\footnote{The same instructions (with a different sequence number) are shown on the screen after blocks 2 and 3. The text in italics and this footnote are not part of the actual instructions.}

\bigskip

This is the end of Sequence 1.

\bigskip

The next sequence is about to begin.

\bigskip

In this sequence, you will belong to the same group of 5 participants as in the previous sequence.

\bigskip

\textit{The following instructions are shown on the screen in COMPETITIVE treatments after the first block is completed.}\footnote{The same instructions (with a different sequence number), with the exception of the sentence in bold, are shown on the screen after blocks 2 and 3. The text in italics and this footnote are not part of the actual instructions.}

\bigskip

This is the end of Sequence 1.

\bigskip

The next sequence is about to begin.

\bigskip

In this sequence, you will belong to the same group of 5 participants as in the previous sequence.

\bigskip

\textbf{Note the change in how your payoff will be calculated.}

\bigskip

You will be ranked in your group based on your score, with rank 1 corresponding to the highest score, and rank 5 to the lowest score. Ties will be broken randomly.

\bigskip

Your payoff will be calculated as follows:

Payoff = \$2.50 $\times$ score, if your rank is 1

Payoff = \$1.50 $\times$ score, if your rank is 2, 3 or 4

Payoff = \$0.50 $\times$ score, if your rank is 5

%%%%%%%%%%%%%%%%%%%%
\newpage
\section{Experimental decision screens}
\label{sec:screens}

\begin{figure}[h!]
\centering
\begin{subfigure}{0.8\textwidth}
   \includegraphics[width=1\linewidth]{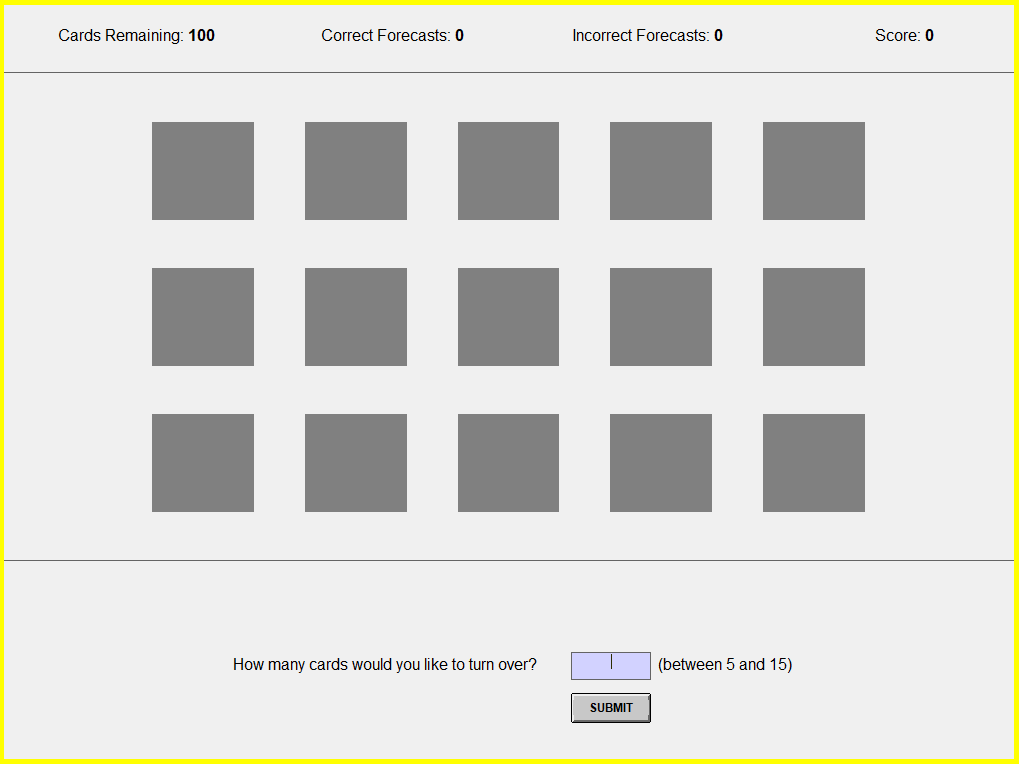}
   \caption*{Card flip decision screen in each period.}
\end{subfigure}
\vskip 1.0cm
\begin{subfigure}{0.8\textwidth}
   \includegraphics[width=1\linewidth]{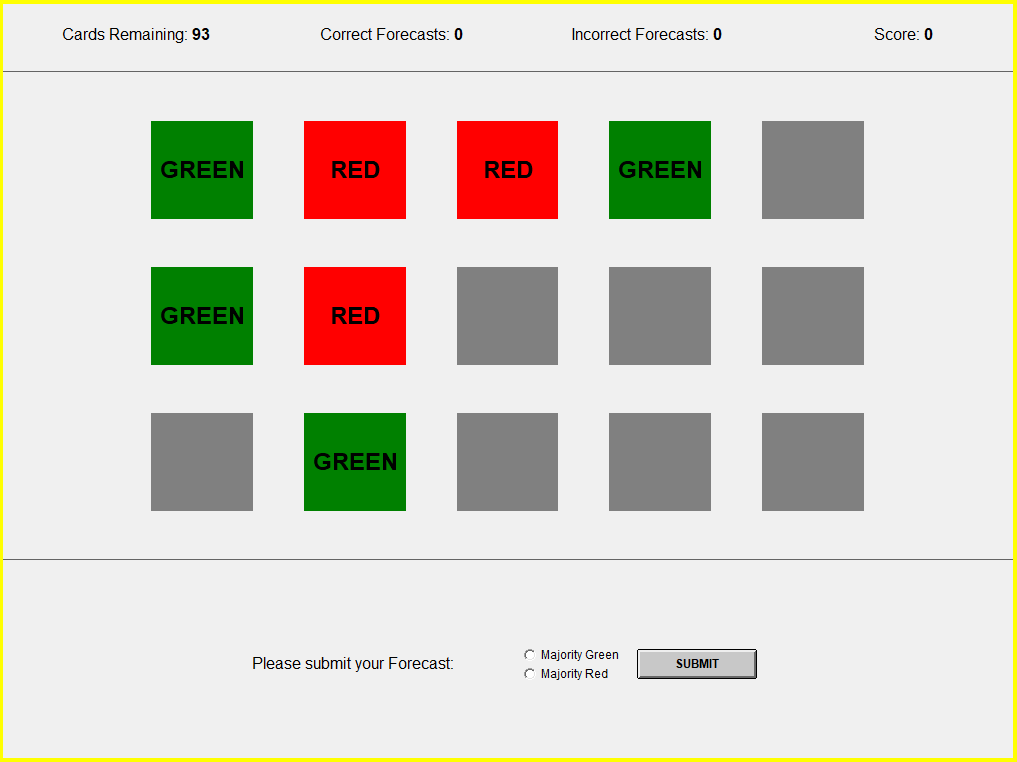}
   \caption*{Majority color forecast screen following subjects' card flip decision.}
\end{subfigure}
%\caption{}
\end{figure}

\begin{figure}[h!]
\centering
\includegraphics[width=0.8\textwidth]{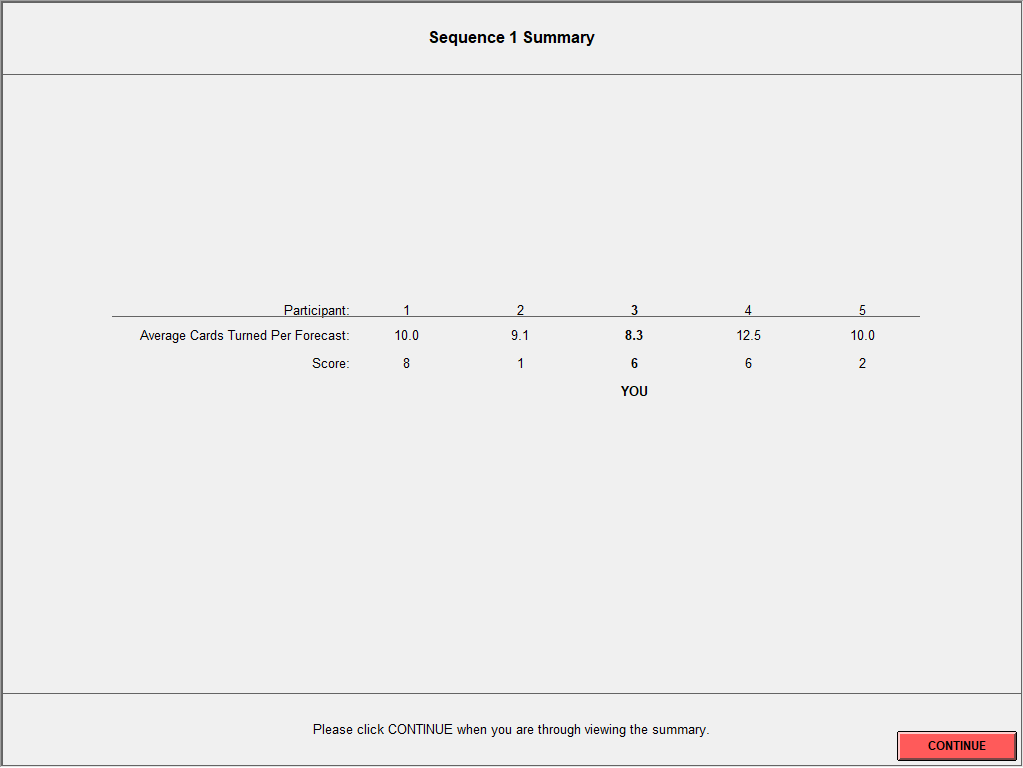}
\caption*{Feedback provided to subjects after each block in the peer information condition including both strategies and scores. In the strategies only treatments, the `Score' row was not present. In the scores only treatment, the `Average Cards Turned Per Forecast' row was not present.}
\end{figure}

\end{document}